\documentclass{aastex631}

\newcommand\degree{\hbox{$^\circ$}}

\begin{document}

\title{Towards a Spectro-Photometric Characterization of the Chilean Night Sky\\ A first quantitative assessment of ALAN across the Coquimbo Region}

\author[0000-0002-0786-7307]{Rodolfo Angeloni}
\affiliation{Gemini Observatory, NSF’s NOIRLab,
Av.~J.~Cisternas 1500 N, 1720236,
La Serena, Chile}

\author[0000-0002-2746-0459]{Juan Pablo Uchima-Tamayo}
\affiliation{Departamento de Astronomía, Universidad de La Serena,
Av.~R.~Bitrán 1305, 1720256,
La Serena, Chile}
\affiliation{Gemini Observatory, NSF’s NOIRLab,
Av.~J.~Cisternas 1500 N, 1720236,
La Serena, Chile}

\author{Marcelo Jaque Arancibia}
\affiliation{Departamento de Astronomía, Universidad de La Serena,
Av.~R.~Bitrán 1305, 1720256,
La Serena, Chile}
\affiliation{Instituto Multidisciplinario de Investigación y Postgrado, Universidad de La Serena,
Av.~R.~Bitrán 1305, 1720256,
La Serena, Chile}

\author{Roque Ruiz-Carmona}
\affiliation{Gemini Observatory, NSF’s NOIRLab,
Av.~J.~Cisternas 1500 N, 1720236,
La Serena, Chile}

\author{Diego Fernández Olivares}
\affiliation{Departamento de Astronomía, Universidad de La Serena,
Av.~R.~Bitrán 1305, 1720256,
La Serena, Chile}

\author{Pedro Sanhueza}
\affiliation{Dirección de Energía, Ciencia y Tecnología e Innovación, Ministerio Relaxiones Exteriores,
Santiago, Chile}

\author{Guillermo Damke}
\affiliation{Cerro Tololo Interamerican Observatory, NSF’s NOIRLab,
Av.~J.~Cisternas 1500 N, 1720236,
La Serena, Chile}

\author{Ricardo Moyano}
\affiliation{Departamento de Astronomía, Universidad de La Serena,
Av.~R.~Bitrán 1305, 1720256,
La Serena, Chile}

\author{Verónica Firpo}
\affiliation{Gemini Observatory, NSF’s NOIRLab,
Av.~J.~Cisternas 1500 N, 1720236,
La Serena, Chile}

\author{Javier Fuentes}
\affiliation{European Southern Observatory,
Av.~Alonso~de~Córdova 3107, 763000,
Santiago de Chile, Chile}

\author{Javier Sayago}
\affiliation{OPCC, NSF’s NOIRLab,
Av.~J.~Cisternas 1500 N, 1720236,
La Serena, Chile}



\begin{abstract}
Light pollution is recognized as a global issue that, like other forms of anthropogenic pollution, has significant impact on ecosystems and adverse effects on living organisms. Multiple evidence suggests that it has been increasing at an unprecedented rate at all spatial scales. Chile, which thanks to its unique environmental conditions has become one of the most prominent astronomical hubs of the world, seems to be no exception. In this paper we present the results of the first observing campaign aimed at quantifying the effects of artificial lights at night (ALAN) on the brightness and colors of Chilean sky. Through the analysis of photometrically calibrated all-sky images captured at four representative sites with an increasing degree of anthropization, and the comparison with state-of-the-art numerical models, we show that significant levels of light pollution have already altered the
appearance of the natural sky even in remote areas. Our observations reveal that the light pollution level recorded in a small town of the Coquimbo Region is comparable with that of Flagstaff, a ten times larger Dark Sky city, and that a mid-size urban area door to the Atacama Desert displays photometric indicators of night sky quality that are typical of the most densely populated regions of Europe. Our results suggest that there is still much to be done in Chile to keep the light pollution phenomenon under control and thus preserve the darkness of its night sky - a natural and cultural heritage that is our responsibility to protect.
\end{abstract}



\section{Introduction\label{sec:intro}}
Light pollution is today acknowledged as a global problem that, similar to other forms of anthropogenic pollution, has significant impacts on ecosystems and wildlife, causing adverse effects on living organisms (\citealt{svechkina2020impact}; \citealt{manriquez2019artificial}; \citealt{irwin2018dark}; \citealt{bennie2015global}; \citealt{chepesiuk2009missing}). 
Originally discussed by the professional astronomical community that was concerned that an inefficient, unnecessary and uncontrolled use of Artificial Light At Night (hereafter ALAN) could dazzle professional observatories by impeding them the vision of the starry sky \citep{walker1970california, riegel1973light}, nowadays ALAN has become a fertile field for interdisciplinary scientific research (e.g., \citealt{cavazzani2022launch, cavazzani2020satellite}; \citealt{kollath2020introducing}; \citealt{jechow2019observing}; \citealt{admiranto2019preliminary}; \citealt{maggi2018trophic}), socio-economic studies (e.g., \citealt{perez2022systematic}; \citealt{gallaway2010economics}; \citealt{nordhaus1994}), and wide-ranging cultural debates (e.g., \citealt{hamacher2020whitening}; \citealt{drake2019our}). 

Similar to other products of human civilization, the effects of ALAN are closely tied to the development and subsequent misuse of more efficient (and often cheaper) energy sources: in this particular case, of the evolving lighting technology \citep{dilaura2008brief}. While the main concern in the 1970s revolved around the transition from mercury-vapor to High-Pressure Sodium (HPS) lamps \citep{riegel1973light}, we are now facing another, potentially more complex transition into the era of Light Emitting Diode (LED) lighting. In fact, the Spectral Power Distribution (SPD) of ALAN is rapidly changing. It has shifted from displaying only a few emission lines to dozens, if not hundreds, of lines, accompanied by a significantly increased continuum level, especially in the blue-green spectral range (see, e.g., Figure  1.14 in \citealt{dutta2017artificial}).

Intimately linked to this technological transition, there is also multiple evidence that light pollution is increasing at an unprecedented rate at virtually all spatial scales (\citealt{kyba2023citizen,kyba2017artificially}; \citealt{sanchez2021first}).
It has been estimated that approximately 99\% of the U.S. and European populations experience light-polluted nights, and more than one-third of the global population has permanently lost the opportunity to observe the Milky Way (hereafter, MW -- \citealt{falchi2016new}). The cultural and social implications of this phenomenon (\citealt{gullberg2019comparison}) are such that some authors have not hesitated to address it as a form of cultural genocide (\citealt{hamacher2020whitening}), thus emphasizing the profound impact it has on our cultural identity and heritage. 

Even remote areas with a reputation for unspoiled nature and pristine skies are not immune to the rapidly evolving effects of light pollution. Chile is renowned worldwide for its diverse and unique ecosystems, and the outstanding quality of its northern skies\footnote{Along with other historical and geopolitical reasons whose discussion goes beyond the scope of this work. The interested reader could refer to \citealt{silva2022astronomo, silva2022atacama, silva2020chile, silva2020stars, silva2019estrellas}.} has transformed it during the last half a century into one of the most important astronomical hubs of the world (\citealt{silva2020chile}).
However, even though there have been longstanding efforts by various public\footnote{The Chilean Ministry of the Environment (MMA) hosts a dedicated section on light pollution in its official website: \texttt{https://luminica.mma.gob.cl/}} and private\footnote{Office for the Protection of the Night Sky Quality of Northern Chile, \textit{a.k.a.} OPCC: \texttt{https://opcc.cl/}} actors to keep the light pollution phenomenon under control, a coordinated program to scientifically characterize, quantify and monitor the extent of its environmental impact over the Chilean sky has only recently begun. Objective and publicly accessible data play in fact a vital role in supporting the scientific evidence that should inform all regulations aimed at mitigating light pollution\footnote{E.g., \texttt{https://opcc.cl/revision\_ds043.html}}, while data-driven indicators of night sky quality are the only reliable way to measure the effectiveness of any technical legislation and to propose focused courses of action \citep{kim2019}.\\ 

Our interdisciplinary Research Group on ALAN was born five years ago with the explicit purpose of filling this need, and aims at providing the broad ALAN community with the first spectro-photometric characterization of one of the most pristine skies of planet Earth. Since 2019 we have been regularly monitoring about twenty distinctive sites (including professional astronomical observatories, astrotourism centers, natural parks, and metropolitan areas) across the Coquimbo Region -- also known as “Región Estrella” (Region of the Stars, in Spanish).
This semi-desert region lies between the Andes Mountains and the Pacific Ocean, a few hundred kilometers north of Santiago. Its high-altitude mountain peaks offer dry and clear skies, making it an ideal host for numerous international astronomical projects, including the southern hemisphere AURA-NOIRLab Programs\footnote{\texttt{https://www.noirlab.edu/public/programs/}}, the Carnegie Institute Las Campanas Observatory\footnote{\texttt{https://www.lco.cl/}}, the ESO La Silla Observatory\footnote{\texttt{https://www.eso.org/public/teles-instr/lasilla/?lang}}, and the future Giant Magellan Telescope\footnote{\texttt{https://giantmagellan.org/}}.
Furthermore, astrotourism\footnote{\texttt{https://astroturismochile.travel/en/home/}}$^,$\footnote{\texttt{https://www.astronomictourism.com/}} has emerged as a fundamental asset of the economic, social, and cultural development of the region. With an abundance of amateur observatories and a wide range of leisure activities related to astronomy, the Coquimbo Region has become a destination that attracts astrotourism enthusiasts from around the world.\\

In this introductory paper we present the results of a first survey program aimed at assessing the impact of ALAN on the natural darkness and colors of four representative sites characterized by an increasing degree of anthropization, and that therefore effectively sample the wide range of average night sky conditions encountered in northern Chile. Section \ref{sec:obs} outlines the criteria that inspired the selection of these specific locations, introduces the instrumental suite, and describes our image acquisition strategy. We also provide a comprehensive description of the data reduction process and analysis.
In Section \ref{sec:results}, we delve into a detailed analysis of the main results obtained at each location. Moving forward to Section \ref{sec:discussion}, we contextualize these findings by comparing the quality of the night sky in the Coquimbo Region with similar sites worldwide, as well as with state-of-the-art numerical models. Finally, in Section \ref{sec:conclusion} we comment on the overall significance, and outline the future steps, of our multidisciplinary endeavor to characterize and preserve the Chilean night sky.

\section{Measurements and methods}\label{sec:obs}
\subsection{Measurement sites}\label{sec:sites}
\begin{figure*}[ht!]
   \centering
    \includegraphics[width=0.9\linewidth]{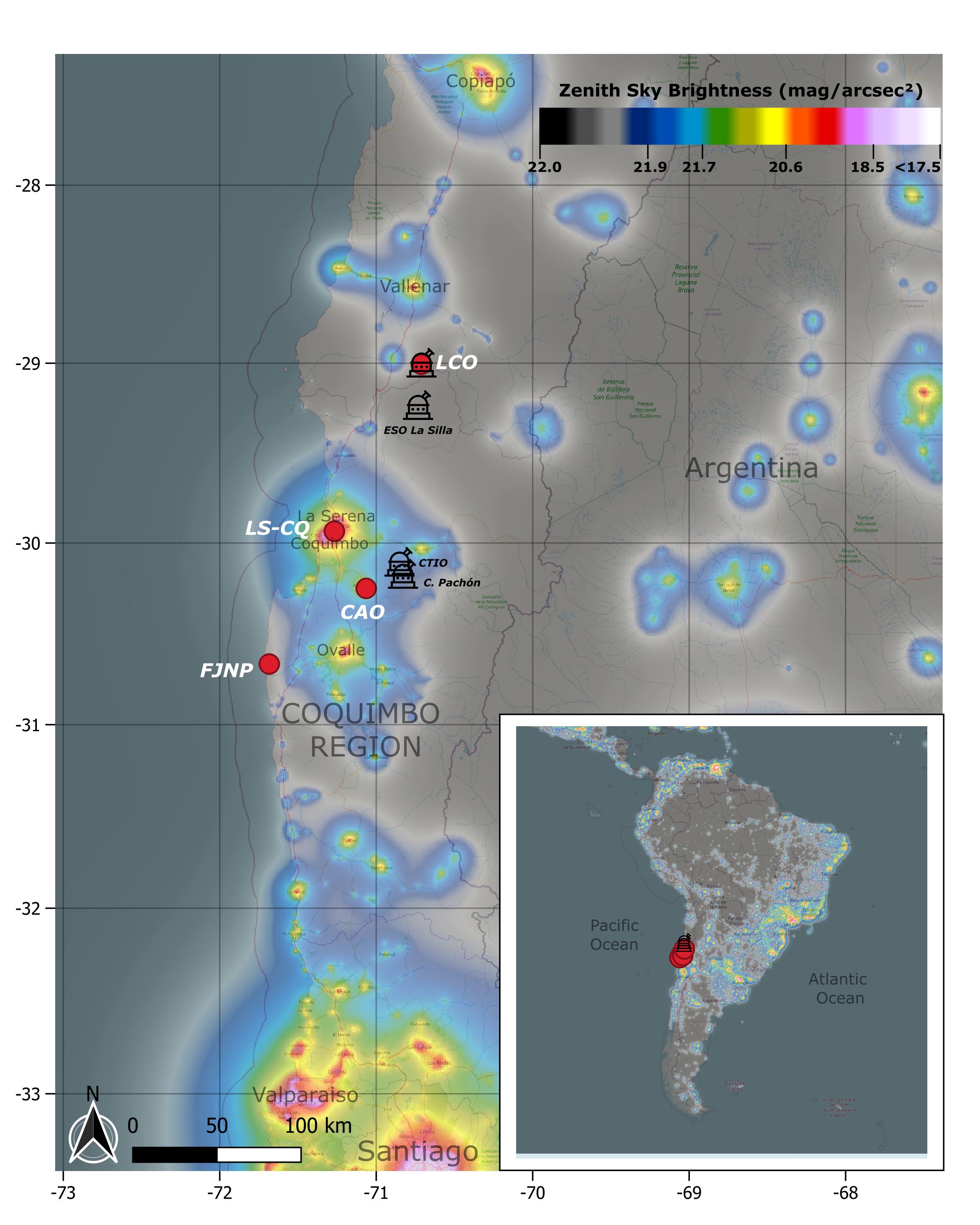}
    \caption{Geographical distribution of the four measurement sites analyzed in this study, represented by red dots with white labels. The dome-shaped icons indicate professional astronomical observatories. Las Campanas Observatory is the one included in the current study. CTIO stands for Cerro Tololo Interamerican Observatory, while Cerro Pachón is home to the Gemini South telescope, SOAR telescope, and V. Rubin Observatory: they will be discussed in a future paper. The background color map is adapted from the New World Atlas of Artificial Night Sky Brightness (\citealt{falchi2016new}).}
    \label{fig:sites}
\end{figure*}

During the site selection process we chose four representative locations that could effectively span the wide range of average night sky conditions expected across northern Chile. In increasing levels of anthropization, they are:
i) Fray Jorge National Park (FJNP), a UNESCO Biosphere Reserve and the first certified Starlight Reserve in South America; ii) Las Campanas Observatory (LCO), a professional astronomical observatory situated on the administrative border between the Atacama and Coquimbo Regions; iii) Collowara Astrotourism Observatory (CAO), an educational and cultural center located close to the mining city of Andacollo; iv) "Greater La Serena" (LS-CQ) that, with a population of 448,784 inhabitants at the 2017 national census, is the fourth largest metropolitan area in Chile. 
Table \ref{tab:infogeo} provides the most relevant geographical information of these four measurement sites, while Figure \ref{fig:sites} shows their geographical distribution across the Coquimbo Region, overlaid on a pseudo-color map adapted from the New World Atlas of Artificial Night Sky Brightness (\citealt{falchi2016new}).

\begin{deluxetable}{ccccccccc}
  \caption{Geographic coordinates obtained from the SQC GPS module of the four measurement sites discussed in this study, listed in order of increasing degree of anthropization. The last four columns indicate the mean distances of each site to the closest urban areas.}
  \label{tab:infogeo}
  \tablehead{
    \colhead{} & \colhead{} & \colhead{} & \colhead{} & \colhead{} & \multicolumn{4}{c}{Distance to} \\ \cline{6-9} 
    \colhead{Measurement site} & \colhead{Acronym} & \colhead{Longitude} & \colhead{Latitude} & \colhead{Elevation} & \colhead{LS-CQ} & \colhead{Ovalle} & \colhead{Vicuña} & \colhead{Vallenar}\\ \colhead{} & \colhead{} & \colhead{\textit{[$\degree \; \arcmin \; \arcsec$]} W} & \colhead{\textit{[$\degree \; \arcmin \; \arcsec$]} S} & \colhead{\textit{[masl]}} 
      & \colhead{\textit{[km]}} & \colhead{\textit{[km]}} & \colhead{\textit{[km]}}  & \colhead{\textit{[km]}}
    }
\startdata
     Fray Jorge National Park & FJNP & 71 39 42.71 & 30 37 35.63 & 278  & 85 & 44 & 112 & 244 \\
     Las Campanas Observatory & LCO & 70 42 01.57 & 29 00 41.98 & 2283 & 117 & 183 & 114 & 49 \\
     Collowara Astroturism Observatory & CAO & 71 03 54.10 & 30 14 56.26 & 1311 & 41 & 41 & 42 & 186 \\
     La Serena - Coquimbo metropolitan area & LS-CQ & 71 16 05.85 & 29 56 04.69 & 15 & -- & 77 & 53 & 155 \\ 
\enddata
\end{deluxetable}

\subsection{Instrumental suite}\label{sec:tools}
Today there is a wide range of tools and technologies available for conducting research on ALAN. Each of these has its own advantages and limitations, as discussed in a few comprehensive reviews as those by \cite{mander2023measure} and \cite{hanel2018measuring} that cover instruments, methods, and interdisciplinary applications. In this subsection, we provide a brief overview of the main features of the Sky Quality Camera (SQC) equipment that we have been using in our monitoring campaign since 2019. For more technical details, we refer the interested reader to the aforementioned reviews and references therein.

Following the classification proposed by \cite{mander2023measure}, SQC can be categorized as a ground-based all-sky imaging sensor. Specifically, the SQCs in our hands are DSLR Canon EOS 6D Mark II cameras, equipped with a 35.8×23.9 mm CMOS sensor of 20.2 (effective) megapixels and coupled with a Sigma EX DG 8 mm f/3.5 fish-eye lens. They provide all-sky multi-spectral information with a spectral coverage and spatial resolution that are not commonly found in ALAN studies. The camera comes with its proprietary software\footnote{Euromix, Ljubljana, Slovenia.} that handles vignetting and distortion corrections of the fish-eye lens and performs astrometric and photometric calibrations. The latter is conducted against the astronomical Johnson-Cousins V band. SQCs directly measure night sky brightness (NSB) values in units of V mag/arcsec$^2$ or, equivalently, luminance in mcd/m$^2$.  The typical photometric errors of SQC data is $\pm$0.01 mag/arcsec$^2$ (Angeloni et al., in preparation).
In addition to NSB measurements, the transformation equations from the camera's RGB channels to the CIE XYZ color space allow us to obtain color correlated temperature (CCT) information \citep{jechow2019noct} with a typical error of $\pm$30 K (Angeloni et al., in preparation).
While there are some known limitations associated with CCT measurements (\citealt{esposito2022correlated}; \citealt{durmus2022correlated}), they still prove valuable when comparing the SPD of the different light sources that contribute to the observed radiance of the night sky \citep{jechow2019dark}.

\subsection{Data acquisition, reduction and analysis}\label{sec:methods}
During the entire surveying campaign, a standardized data acquisition strategy has been implemented to ensure consistency among all observers. In this paragraph, we present a concise overview of this strategy.
The surveying expeditions to the measurement sites are deliberately scheduled during cloudless nights in new Moon periods. In any case, no data is taken when the Moon is at an elevation $\geq$--10$\degree$. The equipment setup takes place before sunset when outdoor light levels are high, allowing for an easier and faster positioning of the SQC\footnote{Mounted on a Manfrotto 055XPro3-3W tripod.} along the north-south, east-west, and zenith axes. This preliminary step is aimed at aligning the sensor focal plane parallel to the horizontal plane, and the camera's optical axis towards zenith. This daytime phase also includes capturing short-exposure frames of the surrounding landscape, that will then aid in accurately tracing the local horizon during the data reduction process. At the nautical twilight, with the sky becoming darker and darker, a series of test images are taken to optimize ISO and exposure time and thus ensure high signal-to-noise ratio while avoiding saturation. With the SQC autodark option enabled, image acquisition can finally start when entering the astronomical twilight.\\

The reduction process, which is performed with the SQC software v.1.9.8, begins with the astrometric calibration of the captured frames. Taking advantage of both the GPS information stored in the RAW data and the software's internal reference star catalog, we align the images following the standard astronomical convention (N up, E to the left and zenith at the center of the azimuthal projection). If the on-site camera setting described in the previous paragraph has been carefully performed, the software corrections are typically minimal ($\leq$ 1$\degree$) on each of the three main axes.
The subsequent crucial step involves tracing the landscape border to accurately define the terrain profile and exclude any foreground obstacles (such as distant mountain tops, trees, or telescope domes) from the active sky area. Before generating the NSB/CCT maps, the contribution of point-source objects is removed and a smoothing function is applied across the entire sky field\footnote{
However, please note that when evaluating SQC vs. GAMBONS NSB maps (see Sec. \ref{subsec:gambons}) we do not remove the point source contribution, nor apply any smoothing. This approach allows for a direct comparison between the observed measurements and the numerical models.}: this step is meant to round off those small, punctual variations in the measured  values that may arise when including very bright objects (such as planets and first magnitudes stars) in the specific area of study.\\

\begin{figure*}
     \centering
    \includegraphics[width=0.45\linewidth]{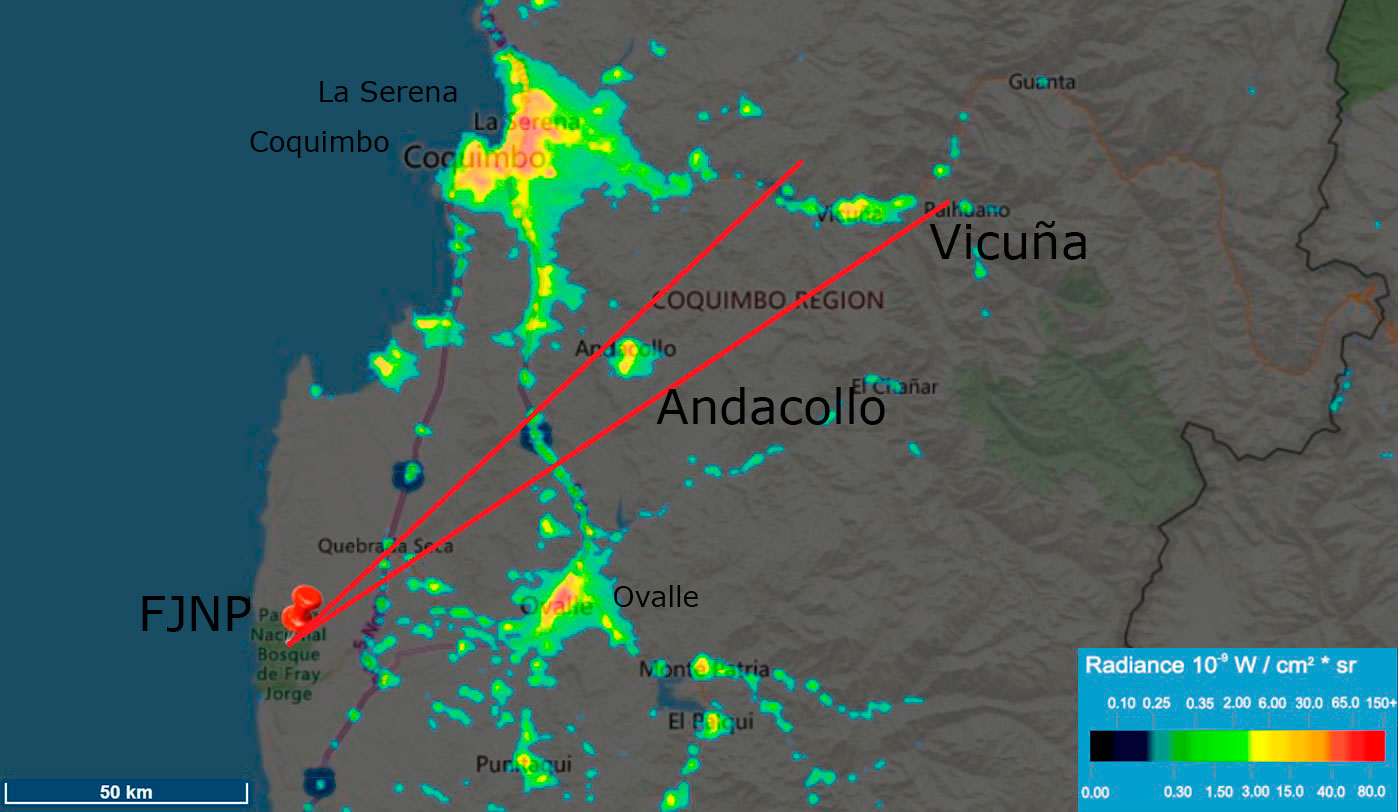}
    \includegraphics[width=0.45\linewidth]{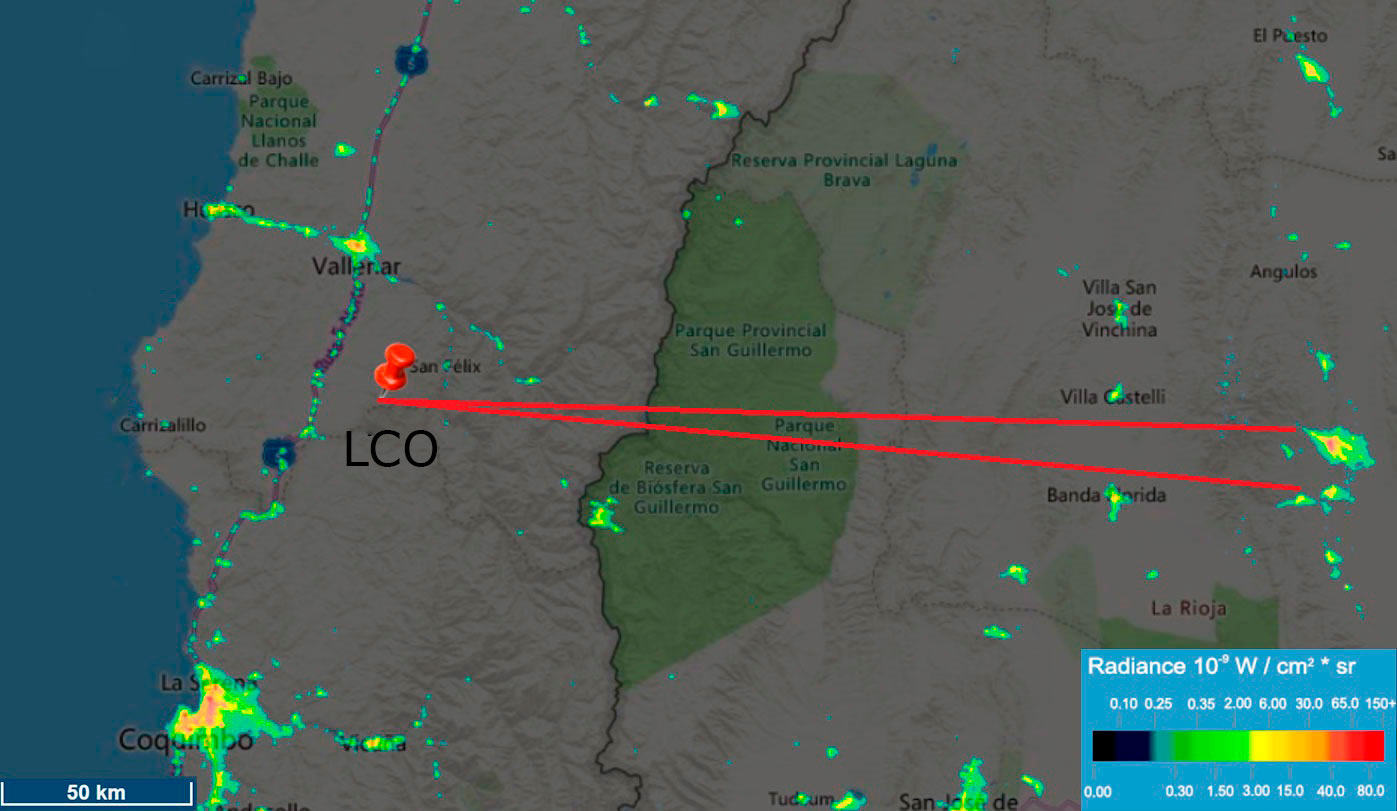}
    \includegraphics[width=0.95\linewidth]{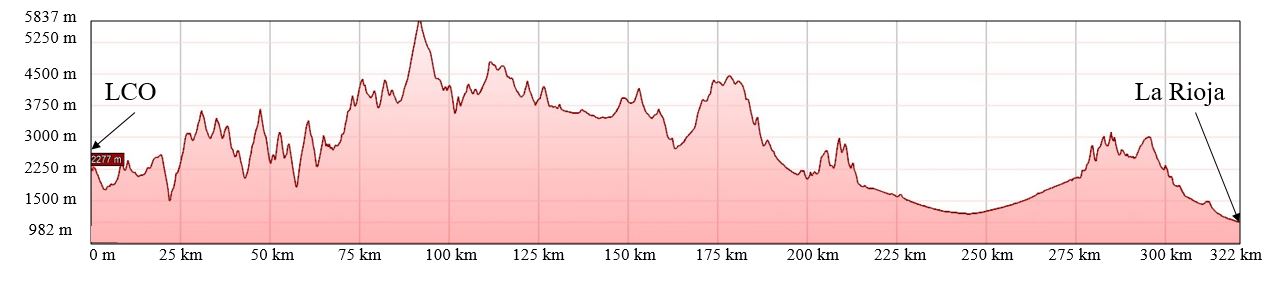}
    \caption{Left panel: Andacollo and Vicuña are located within the same azimuth range when observed from FJNP. Right panel: from LCO there is a specific line of sight towards the east that is devoid of ALAN sources over a distance of several hundred kilometers. These azimuth ranges define the ``ALAN-free'' direction. The 2022 VIIRS/DNB radiance maps adapted from \texttt{http://lightpollutionmap.info} are plotted in the background. Bottom panel: elevation profile along the LCO ALAN-free direction (adapted from Google Earth). See Sect. \ref{sec:obs} and Table \ref{tab:areas} for further methodological details.}
    \label{fig:clean}
\end{figure*}

The initial step in the data analysis involves the visual inspection of both the original RGB and the calibrated NSB/CCT maps to identify any exceeding light contributions across the landscape. NSB/CCT azimuth profiles are then calculated for a series of predetermined elevations (usually, the lowest altitude at which the terrain profile guarantees a 360$\degree$ unobstructed view of the sky,  15$\degree$, and 30$\degree$) to ascertain directions and magnitudes of the various brightness peaks recorded in the images. Using the \textit{lightpollutionmap} web interface\footnote{\texttt{https://www.lightpollutionmap.info}}, we are able to link these local extremes to specific ALAN sources and then determine the corresponding azimuth ranges that encompass their angular extent. Additionally, we exploit the VIIRS/DNB \citep{liao2013suomi} radiance maps plotted over Microsoft Bing base layers to check whether potential ALAN sources visible from space have been recorded in our topocentric SQC data. In certain cases, when observed from a particular vantage point, a specific azimuth range can encompass multiple ALAN bright sources that appear aligned due to the effect of perspective. For example, from FJNP the cities of Andacollo and Vicuña appear to lie along the same line of sight despite the latter is almost two times farther than the former (Figure \ref{fig:clean}, upper left panel). Similarly, the skyglow of Andacollo blends with the light halo of LS-CQ metropolitan area from CAO. \\
In other cases, no ALAN sources are recorded in the VIIRS/DNB radiance maps over large distances. We term these ``dark pathways'' \textit{ALAN-free} directions: their NSB/CCT profiles serve as a baseline to be compared with other lines of sight that are visibly polluted, thus providing a starting point of reference for analysis. For instance, the western horizon of FJNP, facing the Pacific Ocean, serves as a clear example of an ALAN-free direction. Another example is shown in the upper right panel of Figure \ref{fig:clean}: from LCO there is an east-looking line of sight along which no artificial lights are encountered until reaching a few remote villages in the Argentinian La Rioja Province, located over 300 km away behind the 6,000 m Andes wall (Figure \ref{fig:clean}, bottom panel). Their potential contribution to the brightness of LCO sky is negligible, if detectable at all.

Having finally obtained fully calibrated NSB/CCT maps and compiled a list of confirmed and candidate ALAN sources, we can now proceed to measure and compare different regions of the sky to assess the impact of ALAN on each individual site. 

\begin{deluxetable}{ccccccccc}
  \caption{Journal of Observations, showing the local date and time of the expeditions to the sites. Additionally, we report the serial numbers of the SQCs used at each site, the identification numbers of the raw image presented in the study, and the corresponding ISO setting, exposure time, and signal-to-noise ratio.}
  \label{tab:journal-observations}
  \tablehead{\colhead{Site} &  \colhead{Date} & \colhead{Time}  & \colhead{SQC} & \colhead{Frame} & \colhead{ISO} & \colhead{Exp. Time} & \colhead{S/N}\\
      \colhead{\textit{acronym}}   &   \colhead{\textit{yyyy-mm-dd}} & \colhead{\textit{local}} & \colhead{\textit{serial number}} & \colhead{\#} & &  \colhead{\textit{[s]}} & }
  \startdata
    FJNP  & 2021-01-09 & 23:25:50 & 183052000778 & 9872 & 1600 & 120 & 18\\
    LCO  & 2019-11-24 & 02:39:37 & 183052000778 & 9506 & 3200 & 120 & 22\\
    CAO  & 2021-06-10 & 22:41:55 & 423053002003 & 0001 & 1600 & 90 & 30\\
    LS-CQ 1$^{st}$  & 2022-01-02 & 23:25:25 & 423053002003 & 0037 & 1600 & 8 & 39\\
    LS-CQ 2$^{nd}$ & 2022-06-28 & 19:34:42 &423053002002 & 0065 & 1600 & 8 & 42\\ 
  \enddata
 \end{deluxetable}

\begin{deluxetable}{ccccc}
  \caption{ALAN individual sources identified at the four observing sites are grouped in areas of interest, and defined by specific azimuth ranges. The first three azimuth ranges are color-coded and can be recognized in the RGB maps of Figures  \ref{fig:sqc_fjnp}, \ref{fig:sqc_lco}, \ref{fig:sqc_coa}, \ref{fig:sqc_ita}. The corresponding NSB/CCT elevation profiles are plotted in Figures  \ref{fig:vertplot_fjnp}, \ref{fig:vertplot_lco}, \ref{fig:vertplot_cao}, \ref{fig:vertplot_ita}. Please refer to the main text for further details.}
  \label{tab:areas}
  \tablehead{
  \colhead{Site} & \colhead{Area} & \colhead{Azimuth Range} & \colhead{ALAN} & \colhead{Distance} \\
  \colhead{\textit{acronym}} & \colhead{\textit{\#}} & \colhead{[$\degree$]} & \colhead{\textit{Main Sources}} & \colhead{\textit{[km]}}  
  }
\startdata
 {} & {\color[HTML]{3531FF} 1} & 16-36 & LS-CQ & 85,8 \\ \cline{2-5} 
 {} & {\color[HTML]{FE0000} 2} & 78-98 & Ovalle & 44,0 \\ \cline{2-5} 
 {} & {\color[HTML]{34FF34} 3} & 225-315 & \textit{ALAN-free direction} & -- \\ \cline{2-5} 
 FJNP  & 4 & 37-77 & Ruta D-44 & -- \\ \cline{2-5} 
 {} & 5 & 143-145 & Alcones Bajo & 19,0 \\ \cline{2-5} 
 {} & 6 & 50-57 & Andacollo (Vicuña) & 70,0 \\ \cline{2-5} 
 {} & 7 & 150-183 & Gran Valparaíso - RM & 300,0 \\ \hline
 {} & {\color[HTML]{3531FF} 1} & 203-213 & LS-CQ & 117,0 \\ \cline{2-5} 
 {} & {\color[HTML]{FE0000} 2} & 343-2 & Vallenar & 49,0 \\ \cline{2-5} 
 {} & {\color[HTML]{34FF34} 3} & 95-97 & \textit{ALAN-free direction} & -- \\ \cline{2-5} 
 {} & 4 & 260-266 & Ruta 5N @ Cachiyuyo & 19,8 \\ \cline{2-5} 
 {} & 5 & 310-331 & Ruta 5N @ Vizcachitas & 20,0 \\ \cline{2-5} 
 LCO & 6 & 276-292 & Domeyko & 20,3 \\ \cline{2-5} 
 {} & 7 & 238-244 & Ruta 5N @ Observatory exit & 26,0 \\ \cline{2-5} 
 {} & 8 & 34-39 & Alto del Carmen & 35,0 \\ \cline{2-5} 
 {} & 9 & 114-122 & El Veladero open-pit mine & 84,0 \\ \cline{2-5} 
 {} & 10 & 175-182 & Vicuña & 114,0 \\ \cline{2-5} 
 {} & 11 & 193-197 & Ovalle (Andacollo) & 183,0 \\ \cline{2-5} 
 {} & 12 & 7-17 & Copiapó & 186,0 \\ \hline
 {} & {\color[HTML]{3531FF} 1} & {275-5} & Andacollo (275-5) & 2,5 \\ \cline{4-5} 
 {} & {} & {} & LS-CQ (308-345) & 41,0 \\ \cline{2-5} 
 {} & {\color[HTML]{FE0000} 2} & 208-274 & Teck open-pit mine & 2,5 \\ \cline{2-5} 
 CAO & {\color[HTML]{34FF34} 3} & 85-93 & \textit{ALAN-free direction} & - \\ \cline{2-5} 
 {} & 4 & 185-207 & Ovalle & 41,0 \\ \cline{2-5} 
 {} & 5 & 50-61 & Vicuña & 42,0 \\ \cline{2-5} 
 {} & 6 & 168-181 & RM & 350,0 \\ \hline
 {} & {\color[HTML]{3531FF} 1} & {270-360} & Av. del Mar & 1,3 \\ \cline{4-5} 
 {} & {} & {} & Amusement park & 0,5 \\ \cline{2-5} 
 {} & {\color[HTML]{FE0000} 2} & 0-90 & La Serena & - \\ \cline{4-5} 
 {} & {} &  & Cerro Grande & 4,4 \\ \cline{2-5} 
 LS-CQ & {\color[HTML]{34FF34} 3} & 90-180 & San Ramón y Huachalalume & 3,8 \\ \cline{2-5} 
 {} & 4 & 180-270 & Peñuelas - Industrial Distric & 3,0 \\ \cline{4-5} 
 {} & {} & {} & Coquimbo& 8,0 \\ 
\enddata
\end{deluxetable}

\section{Results}\label{sec:results}
In each of the following subsections we provide additional context about each measurement site by highlighting their distinctive characteristics. We then present the results of our survey conducted following the methodology described in Sec. \ref{sec:methods}. The journal of observations is summarized in Table \ref{tab:journal-observations}, while Table \ref{tab:areas} lists the azimuth ranges of interest and the most notable ALAN sources recognizable in the data. It is important to notice that the same color code adopted in Table \ref{tab:areas} for the first three azimuth ranges is maintained in both the topocentric RGB maps (left panels of Figures  \ref{fig:sqc_fjnp}, \ref{fig:sqc_lco}, \ref{fig:sqc_coa}, \ref{fig:sqc_ita}) and in the corresponding NSB/CCT elevation profiles (Figures  \ref{fig:vertplot_fjnp}, \ref{fig:vertplot_lco}, \ref{fig:vertplot_cao}, \ref{fig:vertplot_ita}). These colors identify the areas that include the two brightest ALAN sources (blue and red) and the ALAN-free direction (green) at each site\footnote{With the only exception of LS-CQ, where the very same concept of ALAN-free direction makes no sense anymore -- see Sect. \ref{sec:ls}.}, thus ensuring consistency and readability of the graphical representations.

\subsection{Fray Jorge National Park (FJNP)}\label{sec:frayjorge}
FJNP is located in the Limarí Province of the Coquimbo Region, about 100 km south of LS-CQ. It extends over a total area of 8.863 hectares, of which $\sim$4\% is covered by forests. It is renowned for containing the northernmost Valdivian temperate rain forests: despite being surrounded by semiarid scrublands, in fact, the park's hydrophilic forests have survived for apparently millions of years thanks to the coastal fog\footnote{\textit{Camanchaca} in the Quechua and Aymara language.} which hangs on the mountain-slopes and moistens the subtropical vegetation \citep{cadiz2019phylogeography}. 

The national park was created in 1941 and is today administered by the Chilean forest authority CONAF\footnote{\texttt{https://www.conaf.cl/parques/parque-nacional-bosque-fray-jorge/}.}. In 1977, FJNP joined the UNESCO World Network of Biosphere Reserves. Since 2013, it is also a certified Starlight Reserve (the first in South America and the fourth in the world), namely ``a protected natural area where a commitment to protecting the quality of the night sky and access to starlight is established. Its function is to preserve the quality of the night sky and the different associated values, whether cultural, scientific, astronomical, or the natural landscape''\footnote{According to the definition of Starlight Reserve provided on the official website of the Starlight Foundation: \texttt{https://en.fundacionstarlight.org/contenido/43-definition-starlight-reserves.html}.}.

\begin{figure*}
    \centering
    \includegraphics[width=\linewidth]{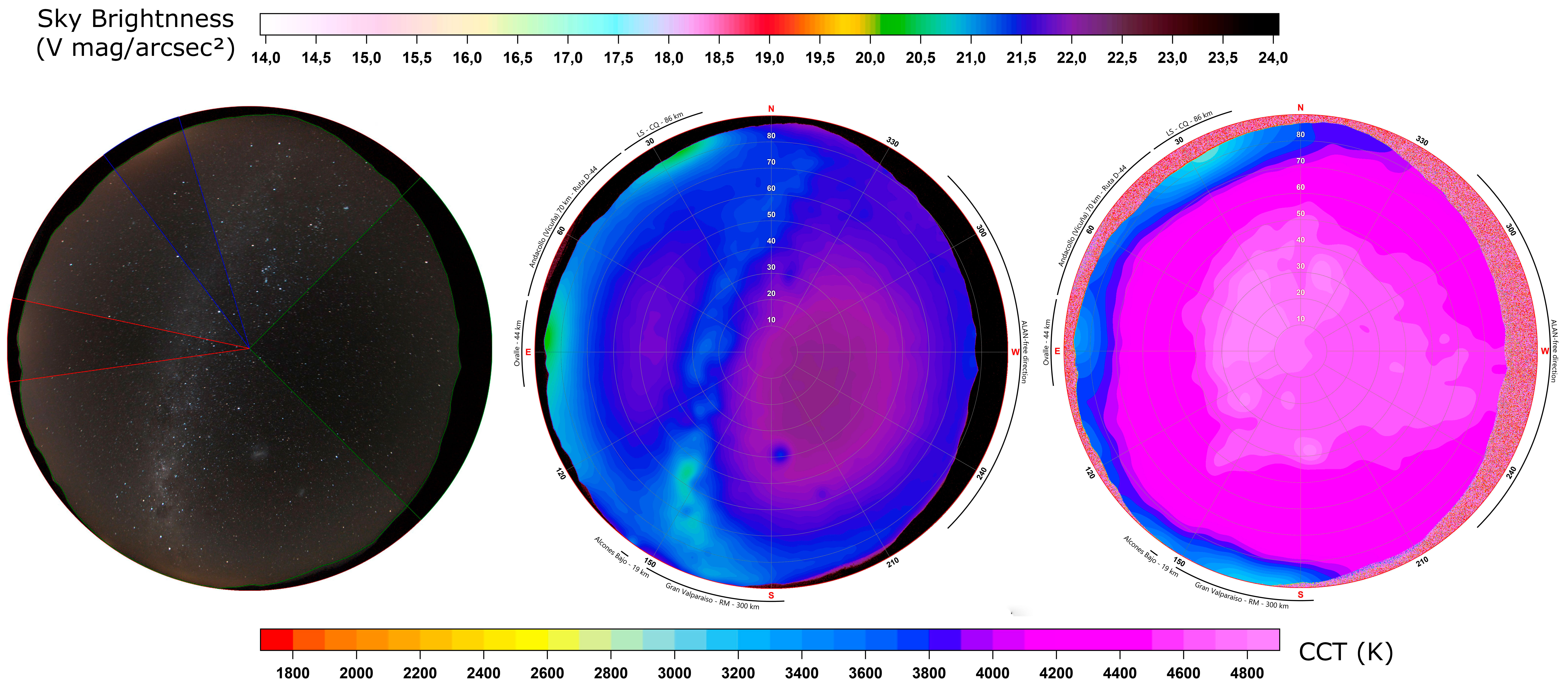}
    \caption{SQC images of the night sky at FJNP captured on 2021 Jan 09. Left panel: RAW RGB frame, also showing as spherical triangles the color-coded azimuth ranges presented in Table \ref{tab:areas} and discussed in Sect. \ref{sec:frayjorge}. Central panel: topocentric NSB map in units of V mag/arcsec$^2$ (upper color bar), highlighting the most prominent ALAN individual sources and their respective distances. Right panel: topocentric CCT map in units of Kelvin (bottom color bar). N is up and E is to the left.}
    \label{fig:sqc_fjnp}
\end{figure*}

\begin{figure*}
    \centering
    \includegraphics[width=0.49\linewidth]{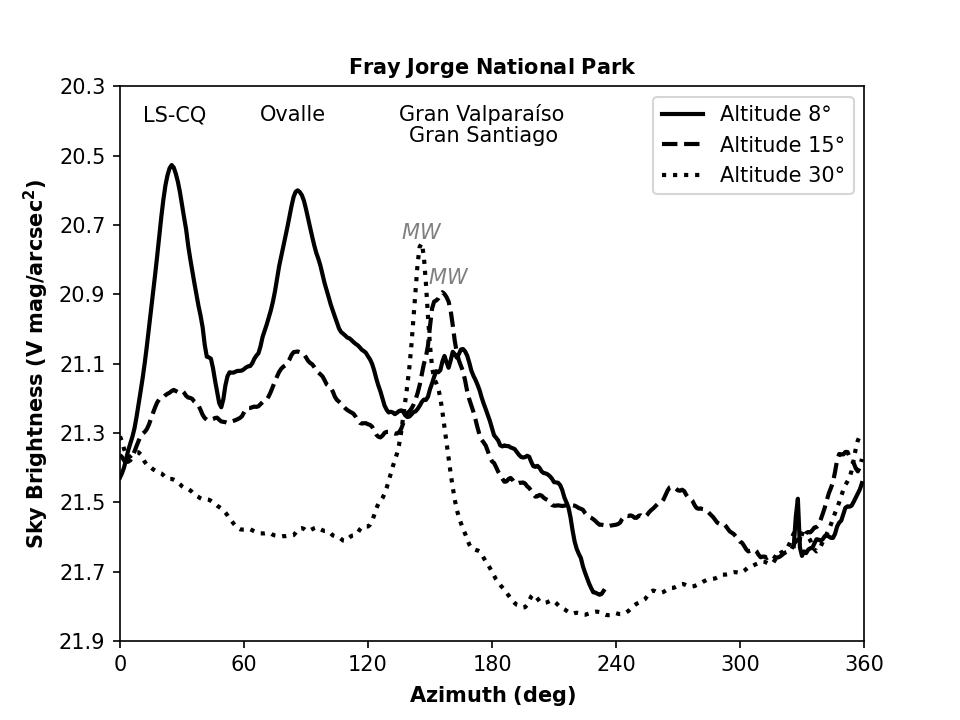}
    \includegraphics[width=0.49\linewidth]{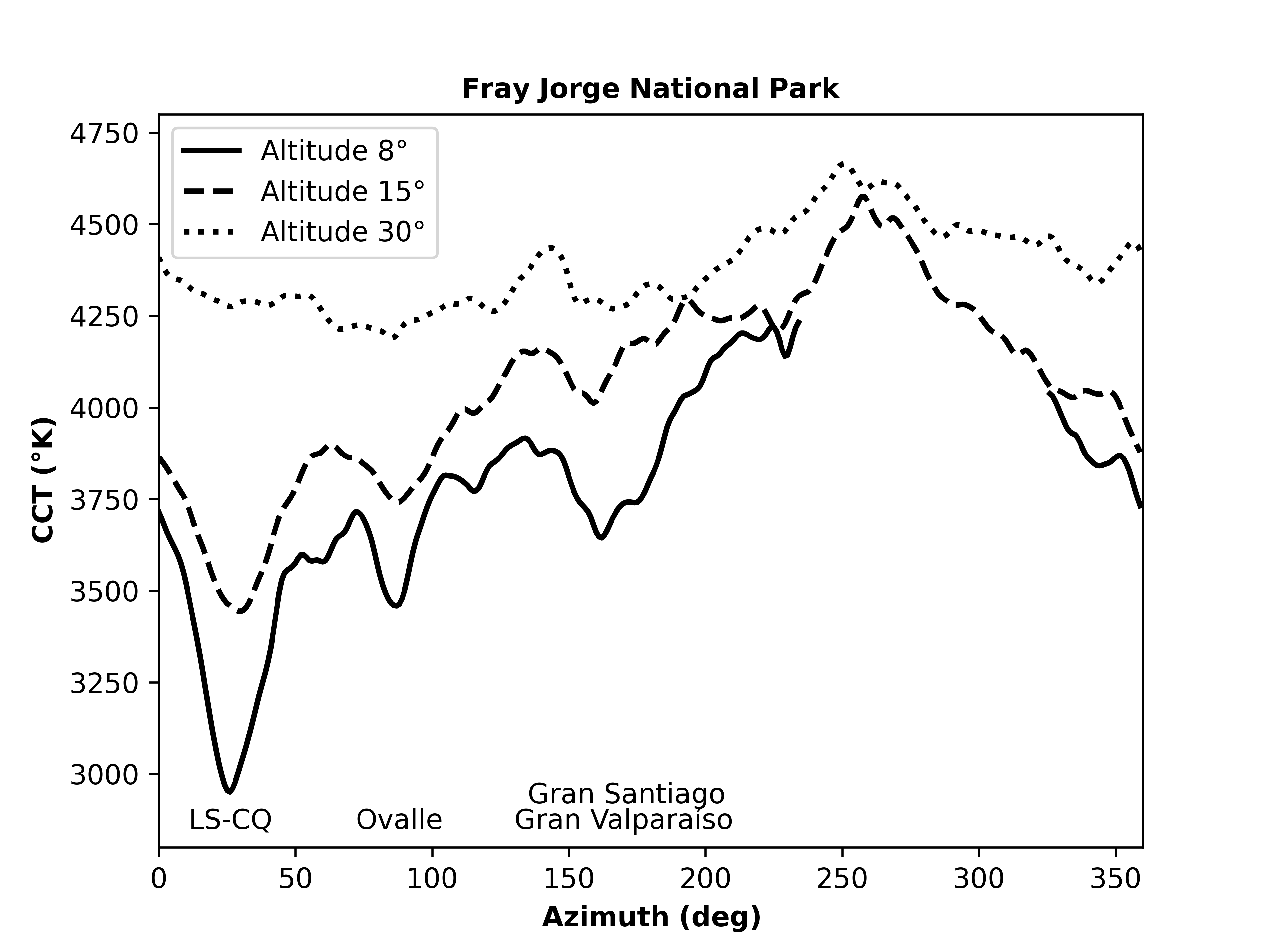}
    \caption{FJNP NSB/CCT profiles averaged within 1$\degree$-wide rings centered at 8$\degree$, 15$\degree$, and 30$\degree$ elevation, as a function of azimuth. Clearly recognizable are the NSB (CCT) local maxima (minima) that indicate the azimuth directions of the strongest ALAN sources. Natural sources (in this case the MW) are indicated with italic gray labels. The typical photometric error of SQC data is comparable to the line thickness. See also Figure \ref{fig:sqc_fjnp} and Table \ref{tab:areas}.}
    \label{fig:fjnp_azring}
\end{figure*}

\begin{figure*}
    \centering
    \includegraphics[width=0.49\linewidth]{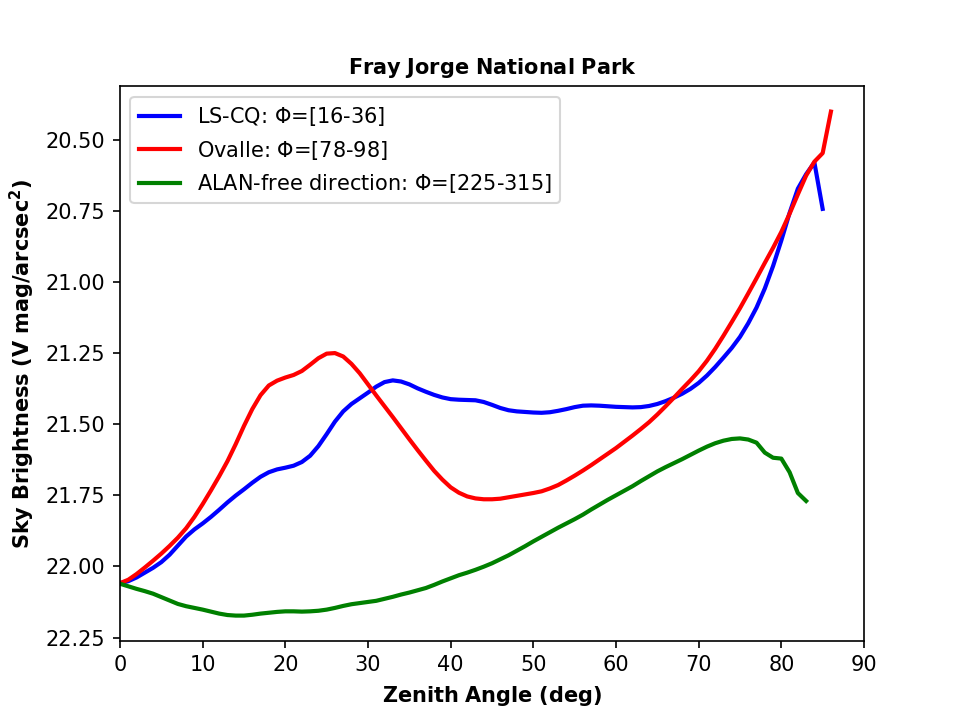}
    \includegraphics[width=0.49\linewidth]{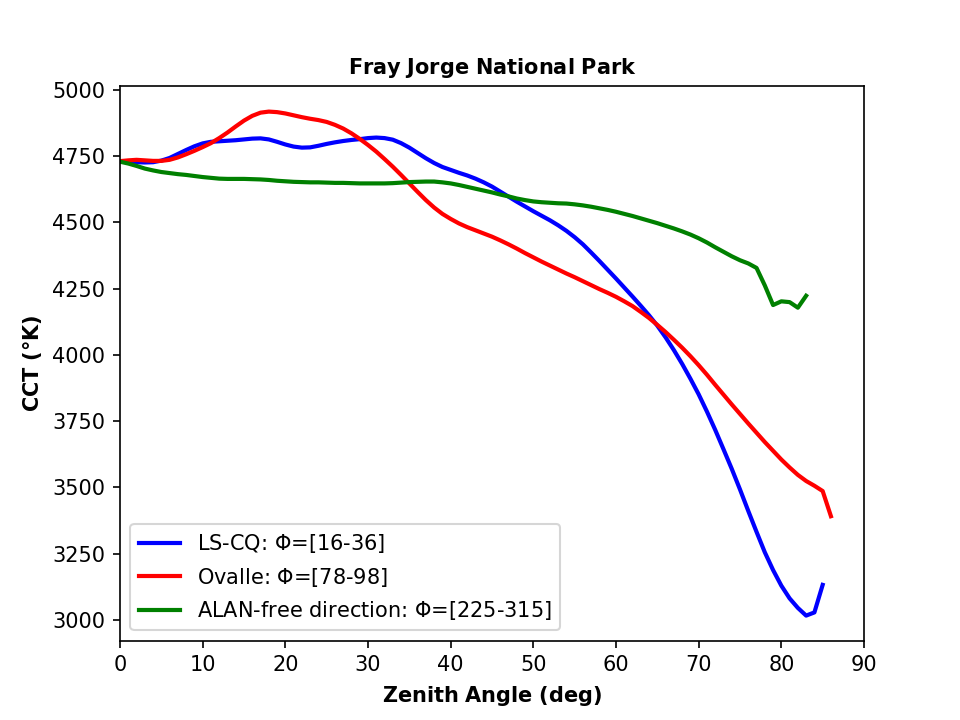}
    \caption{NSB (left panel) and CCT (right panel) profiles as a function of the zenith angle ZA for the three color-coded azimuth ranges identified at FJRP. The typical photometric error of SQC data is also here comparable to the line thickness. For further details refer to Sect. \ref{sec:frayjorge}, Figure  \ref{fig:sqc_fjnp}, and Table \ref{tab:areas}.}
    \label{fig:vertplot_fjnp}
\end{figure*}

Our expedition at FJNP took place during the 2021 southern summer. We chose an observing spot located at an elevation of 278 meters above sea level, approximately 4.5 kilometers away from the coastal line. The natural slope of the terrain causes a small obstruction ($\sim$10\%) of the celestial hemisphere, particularly in the western direction where the lowest $\sim10\degree$ of the sky are hidden by hills.\\

The SQC RGB RAW frame of FJNP (Figure  \ref{fig:sqc_fjnp}, left panel) captures a stunning view of the MW arching high across the sky, accompanied by the presence of the Large Magellanic Cloud (LMC) that has just crossed the southern meridian. The eastern hemisphere, which overlooks the continent, appears already slightly brighter than the western one facing the Pacific Ocean.  By examining the corresponding NSB/CCT maps (Figure  \ref{fig:sqc_fjnp}, middle and right panel), we can clearly identify on the horizon three distinct bright spots of similar intensity. The southern one overlaps with the MW region containing the constellations of Carina, Crux, and Centaurus that have just risen. Figure  \ref{fig:fjnp_azring} allows us to pinpoint these three regions to specific azimuth directions, and in turn to specifically ALAN sources. In order of increasing azimuth and NSB, we recognize: i) the metropolitan area of LS-CQ, with a NSB peak at an azimuth angle AZ$\approx$26$\degree$; ii) the city of Ovalle\footnote{Capital of the Limarí Province.  According to the 2017 national census, Ovalle has a population of 121,269 inhabitants. It is the third largest city in the Coquimbo Region, following Coquimbo and La Serena.}, with a NSB peak at AZ$\approx$86$\degree$; iii) the distant metropolitan areas of Gran Valparaíso\footnote{The third metropolitan area of Chile, with 951,150 inhabitants censored in 2017.} and Gran Santiago, located at least 300 kilometers away, with a NSB peak AZ$\approx$166$\degree$.
The coincidence of azimuth values between the local NSB maxima and CCT minima provides direct evidence of the artificial nature of these light sources. It also exemplifies the valuable role of CCT information in helping differentiate between natural and artificial contributions to the observed radiance.\\

Figure  \ref{fig:vertplot_fjnp} illustrates the variation of NSB/CCT as a function of ZA for the two brightest ALAN sources and the ALAN-free direction (westward quadrant facing the ocean) at FJNP. 
At ZA$\approx$26$\degree$ (ZA$\approx$33$\degree$), towards the direction of Ovalle (LS-CQ), the NSB profiles exhibit local peaks, which can be attributed to the natural modulation of the MK transiting at zenith. As the ZA increases towards the horizon, the NSB virtually parallel profiles of LS-CQ and Ovalle diverge from the ALAN-free profile at ZA$\approx$75$\degree$, suggesting that the western horizon is dominated by ALAN sources up to an elevation of approximately 15$\degree$.
Interestingly, the ALAN-free profile never intersects the other two profiles. The presence of the galactic plane in that direction complicates the interpretation (as the galactic starlight represents one of the most significant contributors to the natural NSB - \citealt{duriscoe2013measuring}; \citealt{masana2021multiband}) but this evidence could be the hint  that even at higher elevation (i.e., for ZA$\geq60\degree$) ALAN may contribute up to  $\sim$0.15 mag/arcsec$^2$ to the NSB of the western horizon at FJNP. When observing under very dark conditions like the ones encountered at this site, the only technique able to confirm this kind of insights seems to be spectroscopy \citep{kollath2023natural}.\\

Based on various measurements and quantitative indicators that include the NSB ratio between the brightest and darkest points in the sky ($\sim$6), the absolute value of the darkest point in the sky (V=22.25 mag/arcsec$^2$) and at zenith (V=22.05 mag/arcsec$^2$), and the dominance of CCT$\geq$4000 K over the celestial hemisphere, it can be confidently stated that the sky at FJNP is still very close to pristine conditions. Its minimal level of light pollution could make it a world-reference location for monitoring the interplay between the various natural components of the night sky radiance without significant interference from artificial sources.

\subsection{Las Campanas Observatory (LCO)}\label{sec:lco}
LCO is a branch of the Astronomy \& Astrophysics division of the Carnegie Institution for Science. It was established in 1969 on a mountain top at an elevation of 2,300 meters in the southern extreme of the Atacama Desert, in visual contact with the ESO La Silla Observatory which is only 26 km away, and approximately 120 km from LS-CQ. Beyond the original 1-m Swope and 2.5-m Du Pont reflectors, LCO now hosts the twin 6.5-m Magellan telescopes and will also be home to the 25-m Giant Magellan Telescope (GMT) currently under construction, one of the first extremely giant telescopes (ELTs) expected to see light in the XXI century.\\

\begin{figure*}
    \centering
    \includegraphics[width=\linewidth]{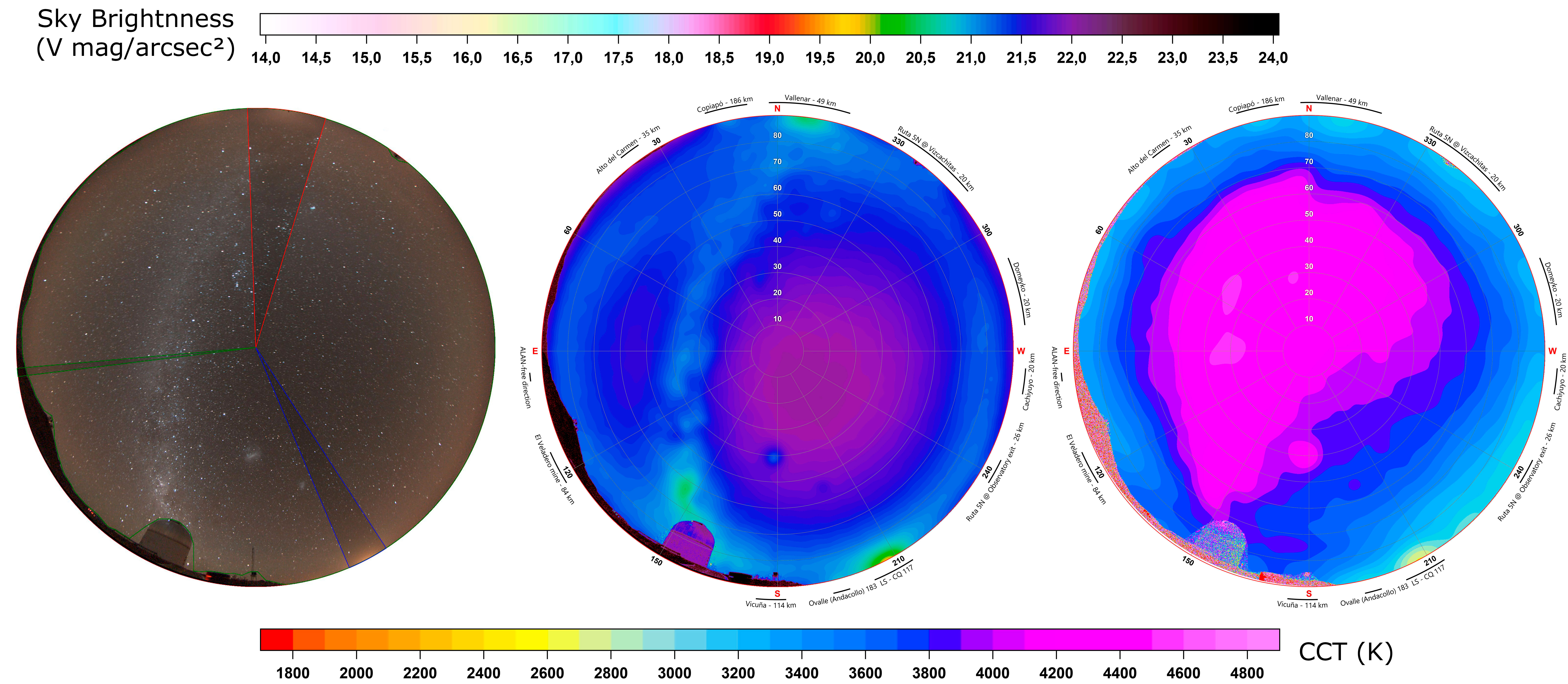}
    \caption{Same as Figure  \ref{fig:sqc_fjnp}, but for LCO on 2019 Nov 24. For further details refer to Sect. \ref{sec:lco} and Table \ref{tab:areas}.}
    \label{fig:sqc_lco}
\end{figure*}

The observing run at LCO took place during the 2019 southern springtime, shortly before the pandemic lockdown that suspended its scientific operations for several months. The reference measurement spot was chosen in the vicinity of the Swope telescope, whose dome is the only relevant obstruction visible towards SE in the RGB frame of Figure  \ref{fig:sqc_lco}, left panel\footnote{The careful observer could locate the Magellan telescope domes on the background hill at AZ$\approx$110$\degree$. The extremely careful observer could even spot the Du Pont telescope dome at AZ$\approx$324$\degree$.}. The observed sky configuration at LCO was similar to that described for FJNP, with the Galactic plane positioned high over the eastern hemisphere and LMC at the meridian transit.

\begin{figure*}
    \centering
    \includegraphics[width=0.49\linewidth]{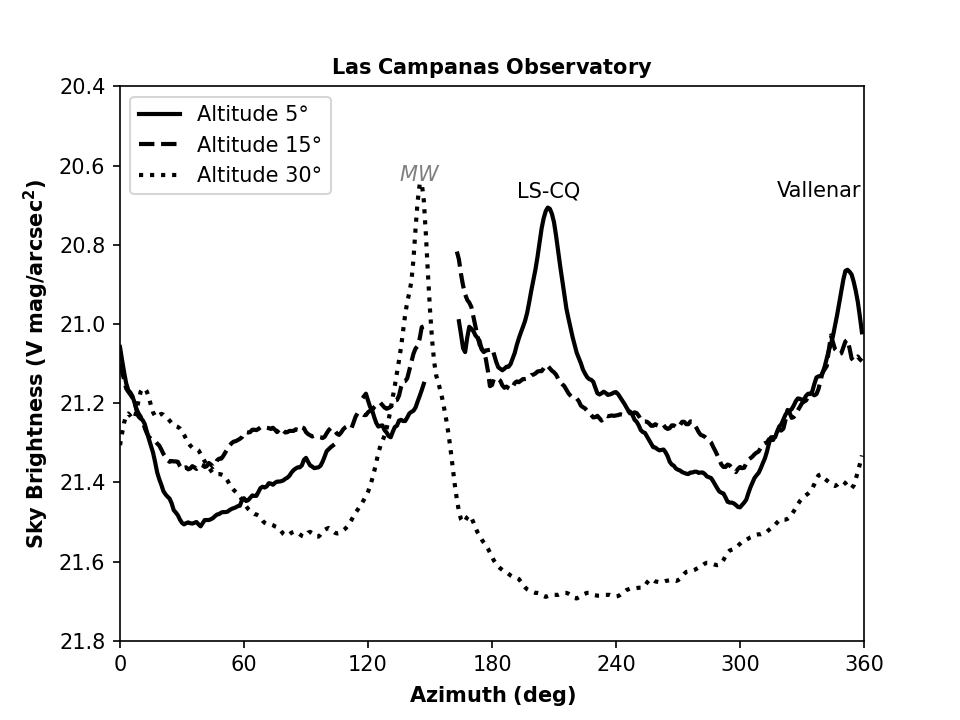}
    \includegraphics[width=0.49\linewidth]{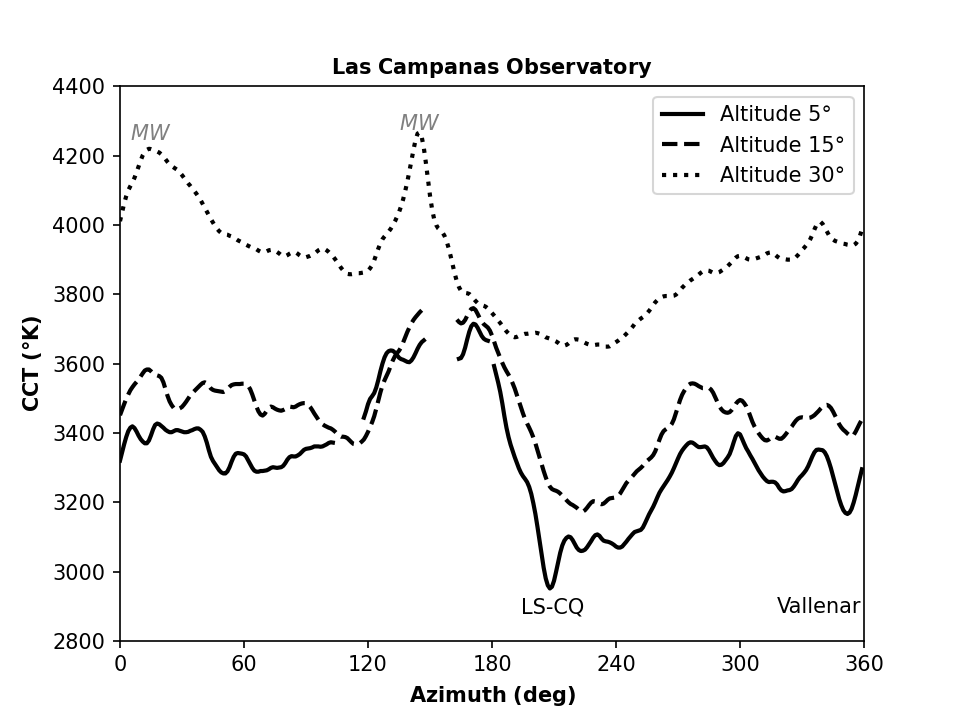}
    \caption{LCO NSB/CCT profiles averaged within 1$\degree$-wide rings centered at 5$\degree$, 15$\degree$, and 30$\degree$ elevation, as a function of azimuth.}
    \label{fig:lco_azring}
\end{figure*}

\begin{figure*}
    \centering
    \includegraphics[width=0.49\linewidth]{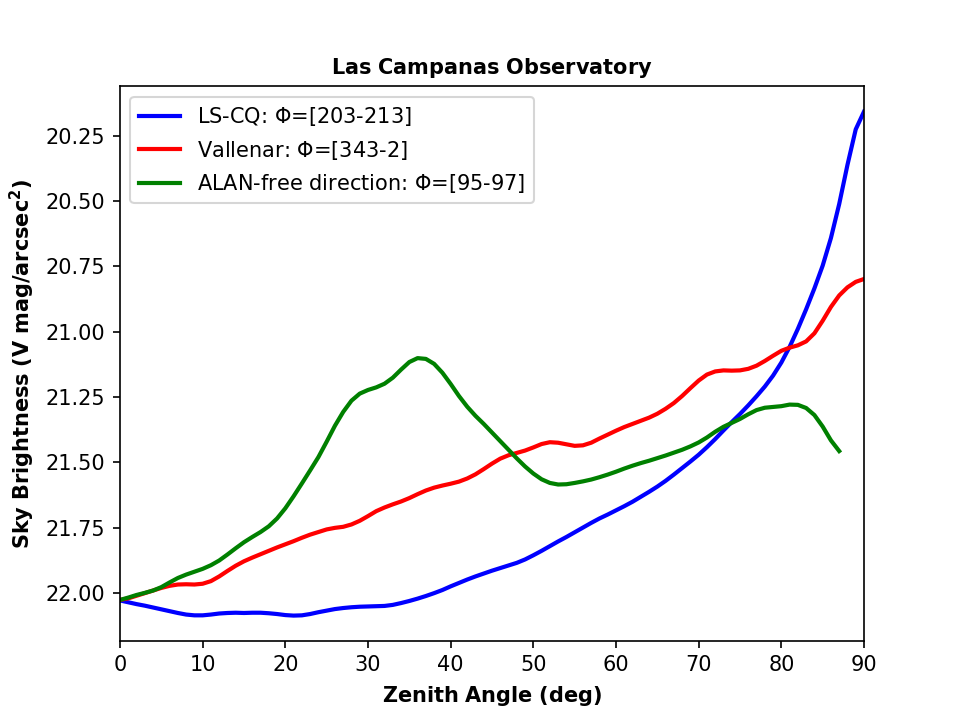}
    \includegraphics[width=0.49\linewidth]{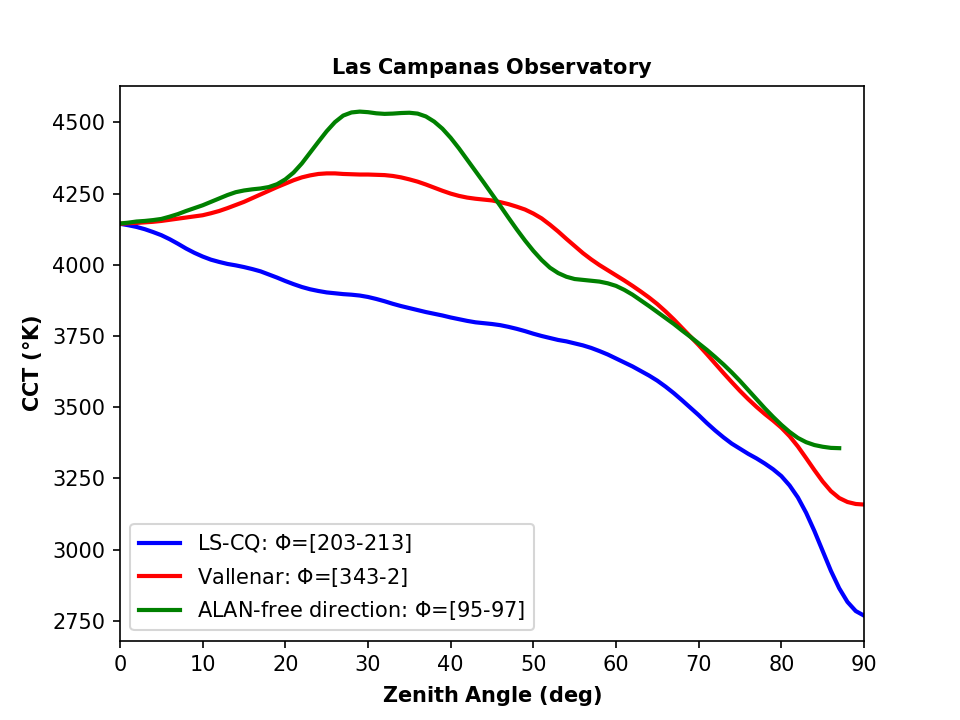}
    \caption{Same as Figure  \ref{fig:vertplot_fjnp}, but for LCO. For further details refer to Sect. \ref{sec:lco}, Figure  \ref{fig:sqc_lco}, and Table \ref{tab:areas}.}
    \label{fig:vertplot_lco}
\end{figure*}

Upon initial examination of the SQC maps in Figure  \ref{fig:sqc_lco} (middle and right panels) and the corresponding azimuth profiles in Figure  \ref{fig:lco_azring}, we can identify two prominent NSB maxima (CCT minima) at approximately AZ$\approx$207$\degree$ and AZ$\approx$352$\degree$, i.e., the skyglow of LS-CQ and Vallenar\footnote{Capital of the Huasco Province in the Atacama Region, with a population of 51,917 inhabitants at the 2017 census.}, respectively. The NSB peak resulting from the rising portion of our Galaxy is here partially obscured by the Swope dome. Figure \ref{fig:vertplot_lco}  illustrates the variations of NSB (left panel) and CCT (right panel) with respect to ZA. The skyglow from LS-CQ dominates the sky up to an elevation of $\sim$15$\degree$ , while the skyglow from Vallenar gradually merges with the MW, resulting in a nearly linear trend in the NSB profile (red line). It becomes more challenging to determine at which elevation the artificial light from the city overwhelms the natural contribution, but a distinctive slope change in both NSB and CCT profiles at ZA$\approx$84$\degree$ suggests that this transition may occur at $\sim$6$\degree$ above the horizon. Along the ALAN-free direction (already discussed in Sec. \ref{sec:methods} for this specific site), the most notable feature is again the modulation visible between $30\degree<ZA<40\degree$ due to the galactic plane. Further examination of Figure  \ref{fig:sqc_lco} reveals additional fainter ALAN sources along the horizon, more easily noticeable in the CCT map. These include the skyglow caps originating from the distant urban areas of Andacollo and Ovalle (aligned along this line of sight, with the latter located approximately 180 km south of LCO) and Copiapó\footnote{Capital of the Atacama Region, with 174,039 inhabitants.} (around 190 km north of LCO).

The CCT map, combined with the georeferenced VIIRS data, also reveals a series of apparently more diluted sources with CCT$\sim$3000 K at the western horizon. They can be ascribed to the (highly debated -- Blanc 2019) renewed lighting system of Ruta 5N, the panamerican highway which connects La Serena with Vallenar and passes about 20 km away from LCO. Four specific directions of interest can be identified: the crossroad between the same highway and the observatory route (AZ$\approx$241$\degree$), and the localities of Cachiyuyo (AZ$\approx$263$\degree$), Domeyko (AZ$\approx$288$\degree$), and Vizcachita (AZ$\approx$320$\degree$). Their contribution to the observed NSB appears minimal even at very low elevation (Figure \ref{fig:lco_azring}): when comparing these directions with their corresponding antipodals, there is no significant increase in NSB levels above $\sim$3$\degree$\footnote{For example, considering a 5$\degree$ reference elevation: at AZ=263$\degree$ (i.e., Cachiyuyo) V=21.32 mag/arcsec$^2$, while at the antipodal AZ=83$\degree$ V=21.38 mag/arcsec$^2$, for a $\Delta$V=0.06 mag/arcsec$^2$.}. However, our SQC data do not allow us to firmly weight the relative contribution of artificial and natural sources: a diffuse light, already perceivable in the RGB frame and clearly visible in the NSB map, runs in fact across the entire horizon and suggests that during that observing night there was a fairly high airglow activity. Airglow is known to contribute significantly to the intrinsic variability of the natural sky (\citealt{kollath2023natural}; \citealt{duriscoe2016photometric}), being its composite and complex multiperiodicity sensitive to both atmospheric and space weather factors \citep{grauer2021linking}. As it was the case for FJNP, under near-pristine conditions like these ones at LCO it becomes challenging to separate the artificial (observed - natural) from the observed (natural + artificial) sky radiance towards specific directions without any spatially resolved spectral information.\\

\subsection{Collowara Astroturism Observatory (CAO)}\label{sec:collowara}
CAO is an astrotourism and educational complex situated at an altitude of 1,300 masl on Cerro El Churqui, approximately 2 km away from both the city of Andacollo\footnote{Population of 11,044 inhabitants at the 2017 census.} and the open-pit mine TECK ``Carmen de Andacollo''. It is positioned at the center of a hypothetical triangle, equidistant (around 40 km) from LS-CQ, Ovalle, and Vicuña. Managed by the local lodging network, CAO was officially inaugurated on June 25th, 2004, and has been welcoming tourists of all ages and nationalities year-round ever since (with the notable exception of the pandemic period, during which our study visit was nevertheless authorized). 
During our visit to CAO in the winter of 2021, we installed the SQC on a plateau slightly higher than the main observatory building. This vantage point allowed us to gain an unobstructed 360$\degree$ view of the local horizon and a direct line of sight over the entire city and mine areas. \\

\begin{figure*}
    \centering
    \includegraphics[width=\linewidth]{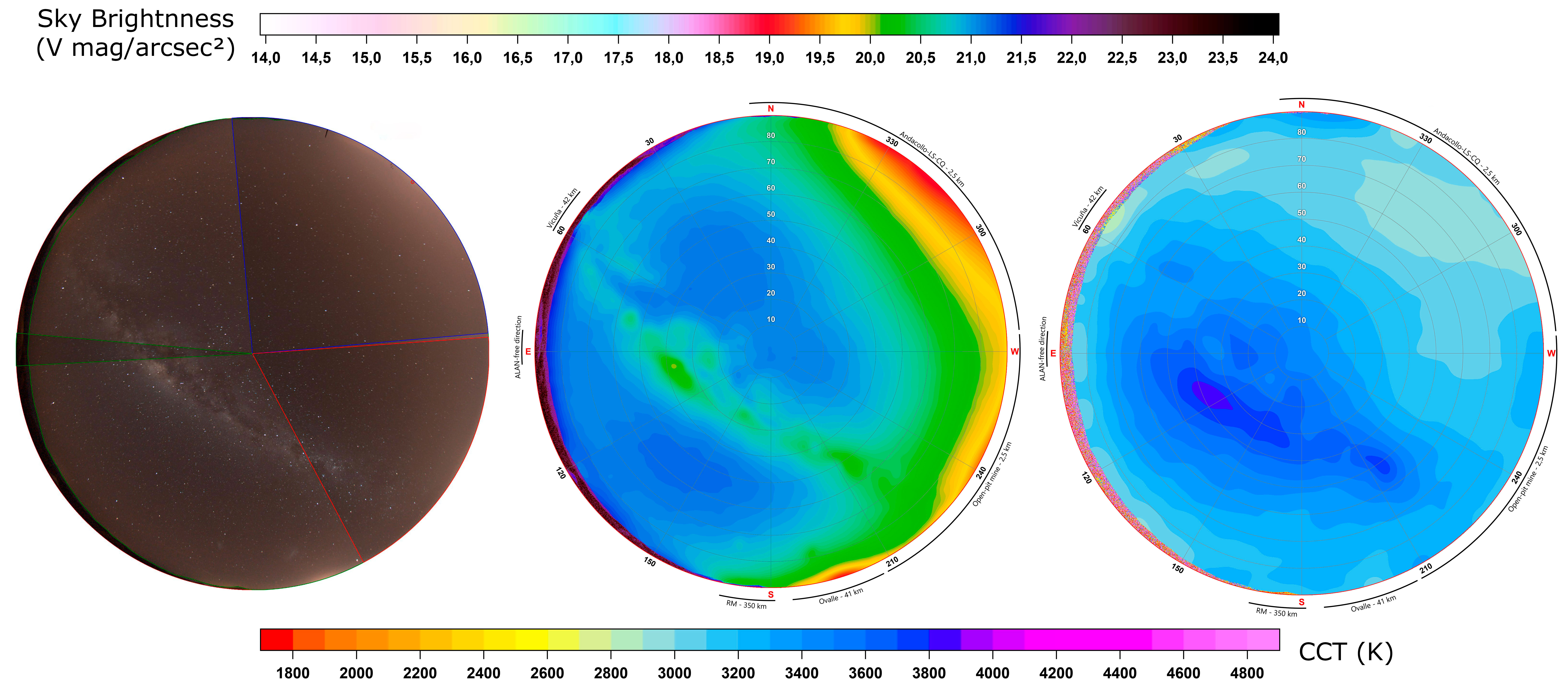}
    \caption{Same as Figure  \ref{fig:sqc_fjnp}, but for CAO on 2021 Jun 10. For further details refer to Sect. \ref{sec:collowara} and Table \ref{tab:areas}.}
    \label{fig:sqc_coa}
\end{figure*}

\begin{figure*}
    \centering
    \includegraphics[width=0.49\linewidth]{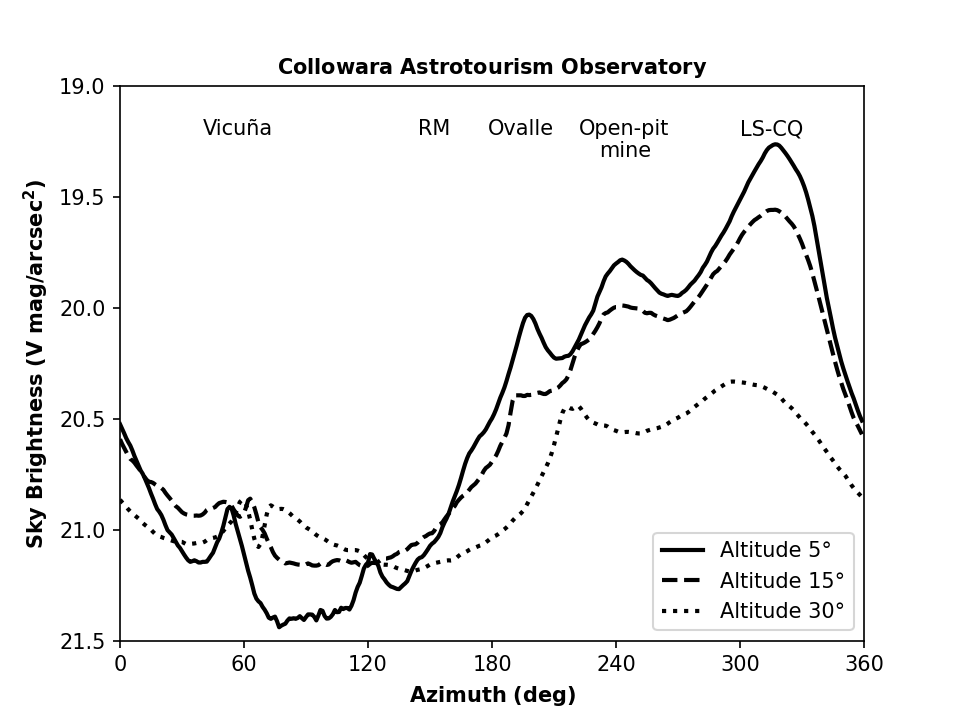}
    \includegraphics[width=0.49\linewidth]{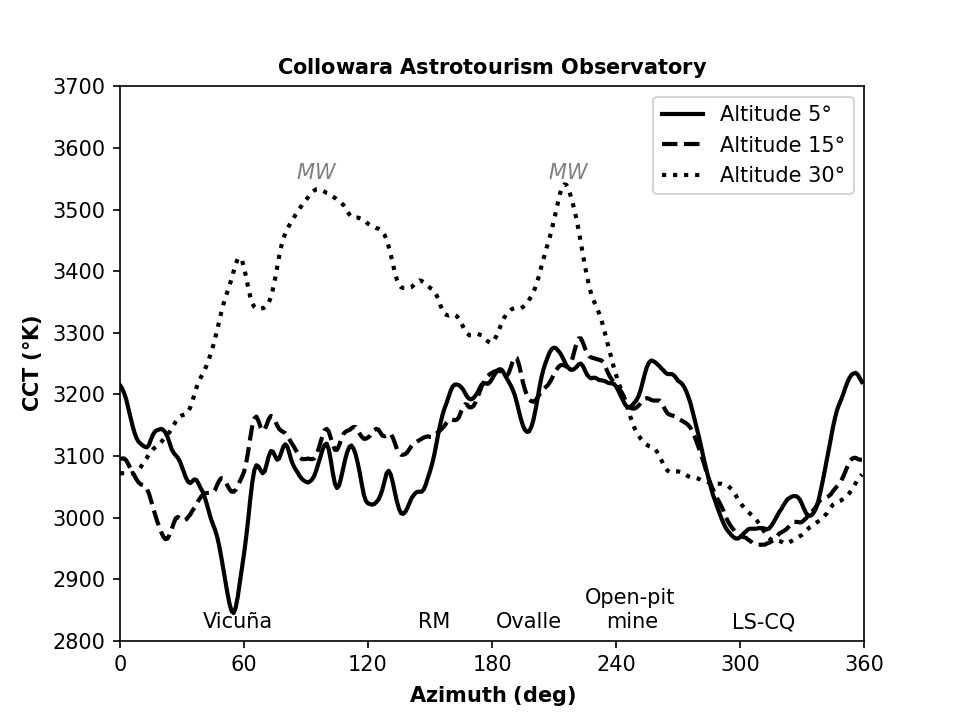}
    \caption{CAO NSB/CCT profiles averaged within 1 $\degree$-wide rings centered at 5$\degree$ elevation, 15$\degree$, and 30$\degree$, as a function of azimuth (see also Table \ref{tab:areas}).}
    \label{fig:cao_azring}
\end{figure*}

In the winter and at these southern latitudes, the bulge and inner disk of our Galaxy reach their highest point above the horizon. They constitute by far the strongest natural contribution of the zenith sky radiance during Moon-less nights (see, e.g., Figure  4 in \citealt{duriscoe2016photometric}). But the RGB image of CAO shown in Figure  \ref{fig:sqc_coa} immediately tells us that the impact of light pollution  at this site can no longer be considered negligible. The celestial dome appears divided into two almost equal sections. Because of a higher sky background, an apparently discolored MW is perceptible towards the east, still showcasing its dark dust lanes that have long fascinated the local ancestors in pre-ALAN eras (\citealt{gullberg2019comparison}). But, on the other side of the celestial vault, three equally spaced ALAN caps dazzle the entire family of setting constellations.
The NSB/CCT maps in Figure  \ref{fig:sqc_coa}, along with their corresponding azimuth profiles in Figure  \ref{fig:cao_azring}, confirm that the largest and brightest skyglow, with an NSB peak at AZ$\approx$317$\degree$, is the merged ALAN of LS-CQ and Andacollo. The other two skyglows can be linked to the nearby open-pit mine and to the city of Ovalle, this latter located sixteen times farther away (with NSB peaks at AZ$\approx$244$\degree$ and AZ$\approx$198$\degree$, respectively). The skyglow from Ovalle also exhibits pretty symmetric wings, with the southern one being the contribution of the even more distant Gran Valparaíso (approximately 310 km away) within the azimuth range 183$\degree$-190$\degree$, and the southwestern one arising from the D-43 street connecting Ovalle and LS-CQ. Additionally, our data detects the skyglow from the Gran Santiago metropolitan area (approximately 350 km away!) towards AZ$\approx$174$\degree$.

The CCT map provides a clearer identification of Vicuña (AZ$\approx$53$\degree$) and seems to have also recorded the ALAN effects of the Argentinian cities of San Juan ($\sim$280 km away) and Mendoza ($\sim$360 km away), as indicated by a CCT  minimum of $\sim$3000 K spread over the azimuth range 115$\degree$-152$\degree$. While the proposed interpretation may sound questionable for the presence of the physical barrier embodied by the Andes, it is worth noticing that satellite meteorological data\footnote{Available from, e.g., \texttt{https://zoom.earth/}} reported the presence of high clouds passing over west Argentina during the night of our observation. These scattered cloud coverage may have amplified the effects of urban light pollution (\citealt{kyba2011cloud}; \citealt{jechow2017imaging}), making them detectable from the Chilean side of the \textit{Cordillera} up to the moderate altitude of CAO. However, this hypothesis remains highly speculative, and additional measurements under all possible weather conditions, preferably in the form of continuous all-sky monitoring campaigns, are highly encouraged.\\

\begin{figure*}
    \centering
    \includegraphics[width=0.49\linewidth]{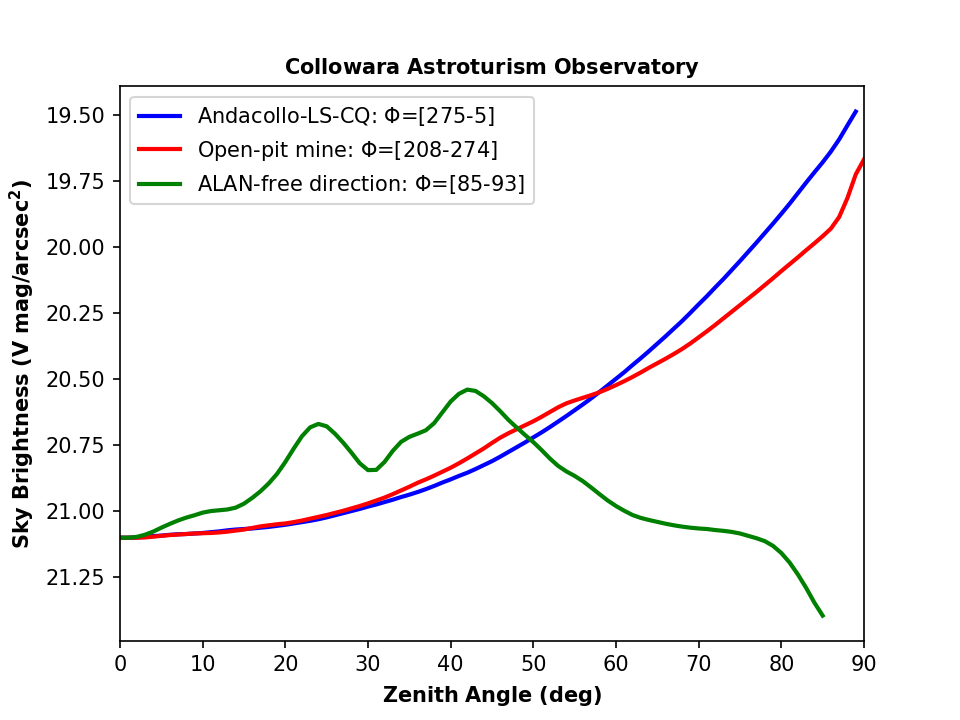}
    \includegraphics[width=0.49\linewidth]{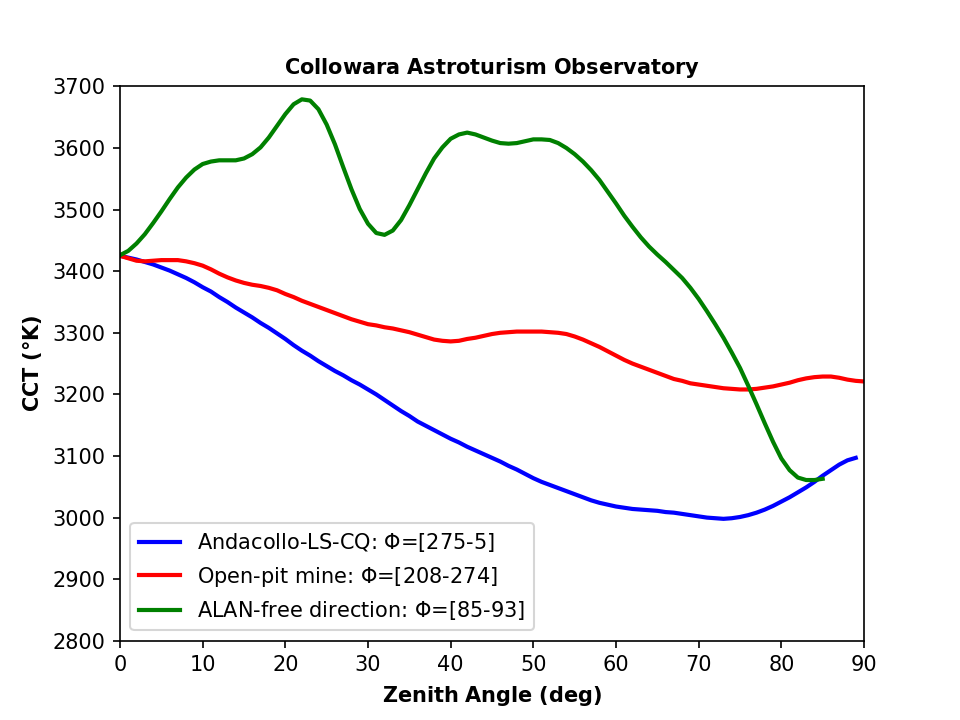}
    \caption{Same as Figure  \ref{fig:vertplot_fjnp}, but for CAO. For further details refer to Sect. \ref{sec:collowara}, Figure  \ref{fig:sqc_coa}, and Table \ref{tab:areas}.}
    \label{fig:vertplot_cao}
\end{figure*}

The NSB/CCT profiles graphed in Figure  \ref{fig:vertplot_cao} confirm that light pollution at CAO makes the western horizon approximately seven times brighter (at a reference elevation of 5$\degree$) compared to the chosen ALAN-free direction. 
Light pollution at CAO has not only canceled out the majority of $\geq$5 mag stars, but it has already lightened zenith (the entire sky) by 54\% (130\%) over the natural value, invariably altering the night sky quality indicators shown in Table \ref{tab:minmax}.

\subsection{La Serena - Coquimbo metropolitan area (LS-CQ)}\label{sec:ls}
La Serena and Coquimbo are neighboring cities that together shape the fourth largest metropolitan area in Chile. Located approximately 470 km north of Santiago, it serves as a gateway to the Atacama Desert. Situated along the Pacific Ocean, these two cities are renowned for their historical charm and cultural significance. La Serena, the regional capital, is also the second oldest city in Chile after the capital Santiago. For the past half a century, it has been home to the Chilean headquarters of AURA-NOIRLab and LCO. In recent years, the entire metropolitan area has witnessed rapid and somewhat uncontrolled urban and demographic expansion \citep{orellana2020conformacion}. 

Our reference spot to study the night sky quality in the urban fabric of LS-CQ became the football field of the local ``A. De Gasperi'' Italian School. Despite its relative proximity to both the Ruta 5N panamerican highway and the bustling tourist promenade of ``Avenida del Mar'', this didactic compound still provided a relatively large surface with minimally intrusive lighting installations. In 2021 we visited the Italian School twice, first in January during the summer tourist season and then in June at the beginning of the local winter. Because of the recent opening of an overly illuminated sports ground in close proximity, today this monitoring site is no longer suitable for conducting NSB measurements - already an indirect hint of the rapidly increasing level of light pollution in the city.\\

\begin{figure*}
    \centering
    \includegraphics[width=\linewidth]{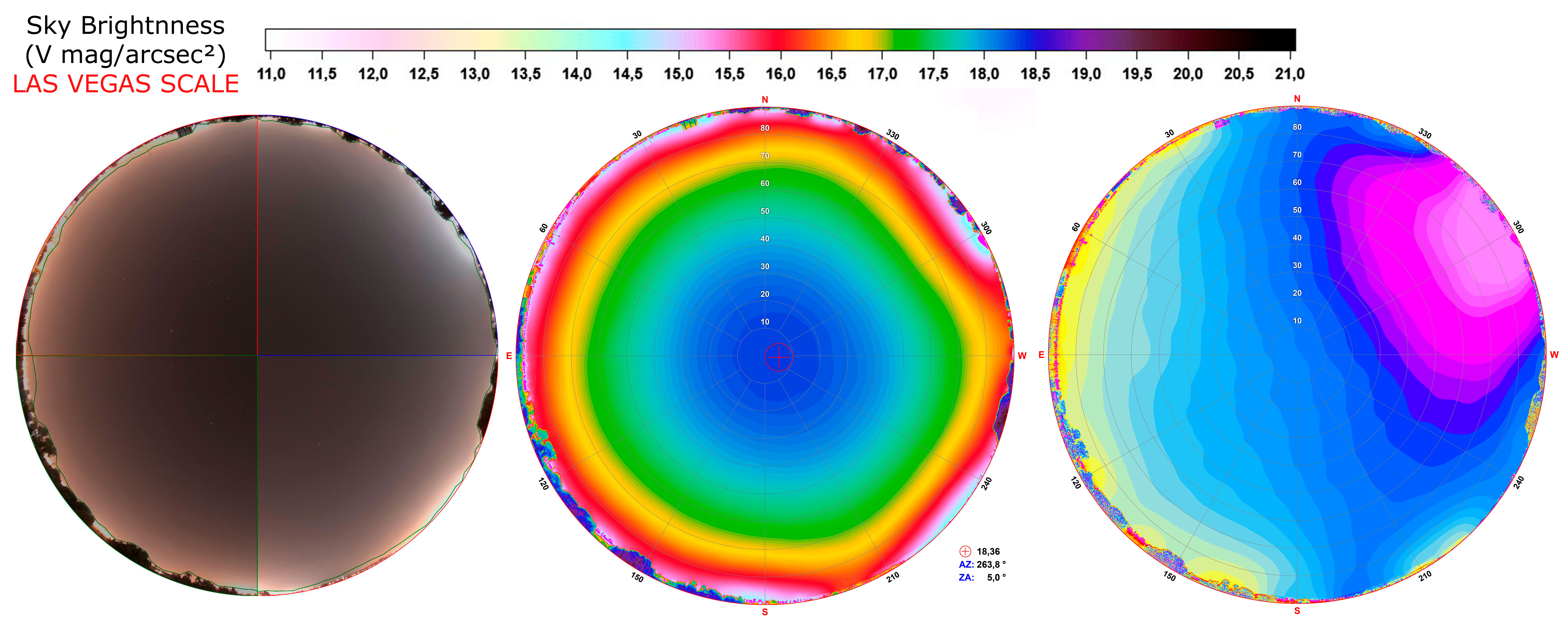}
     \includegraphics[width=\linewidth]{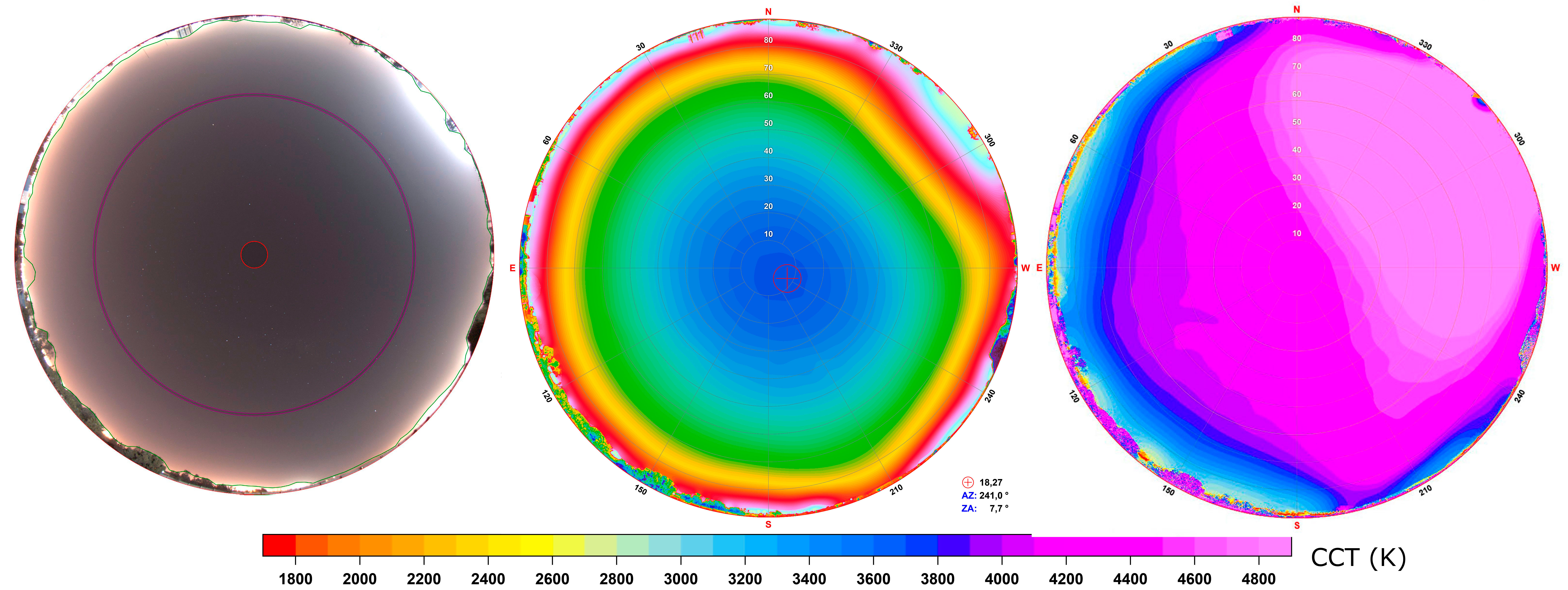}
    \caption{As in Figure  \ref{fig:sqc_fjnp}, but for LS-CQ on 2022 Jan 2 (upper row) and Jun 28 (bottom row). In the second epoch RGB image (bottom left), the 5$\degree$ zenith circle and the 30$\degree$ elevation ring which are mentioned in the text and described in Table \ref{tab:minmax} are explicitly marked. It is important to note the different color scale used in the NSB maps, that needs to accommodate the shifted range of brightness levels in this heavily light-polluted urban sky. Additionally, there is a significant presence of areas with CCT well over 4000 K. Unlike the case discussed for FJNP in Section \ref{sec:frayjorge}, here we are witnessing the invasion of white LED sources, particularly from private buildings and commercial establishments. For further details refer to Sects. \ref{sec:ls} and \ref{sec:discussion}.}
    \label{fig:sqc_ita}
\end{figure*}

\begin{figure*}
    \centering
    \includegraphics[width=0.49\linewidth]{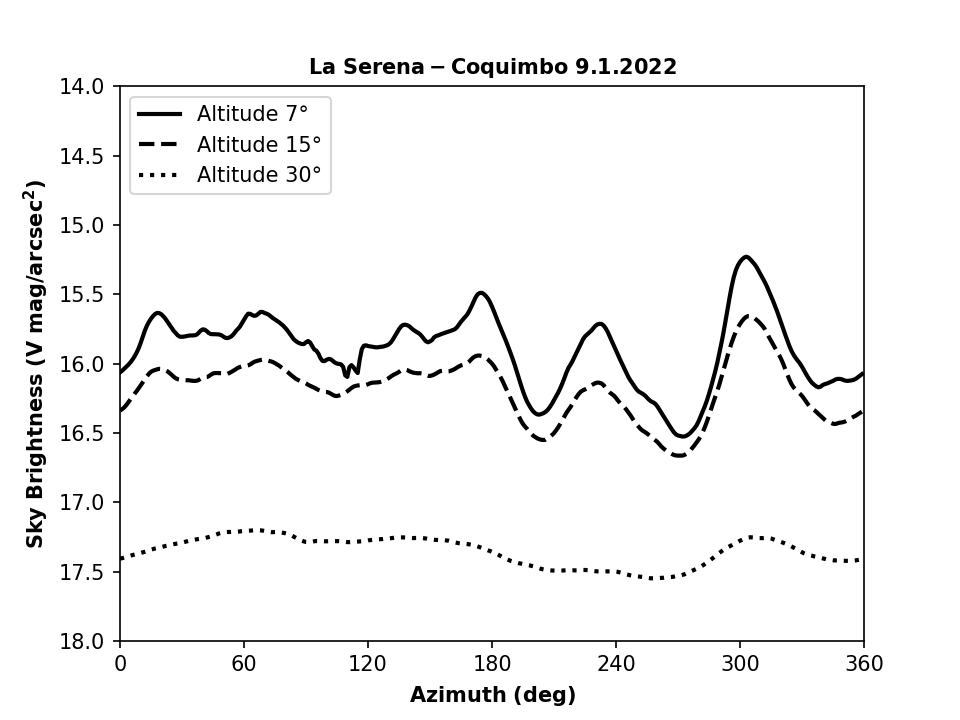}
    \includegraphics[width=0.49\linewidth]{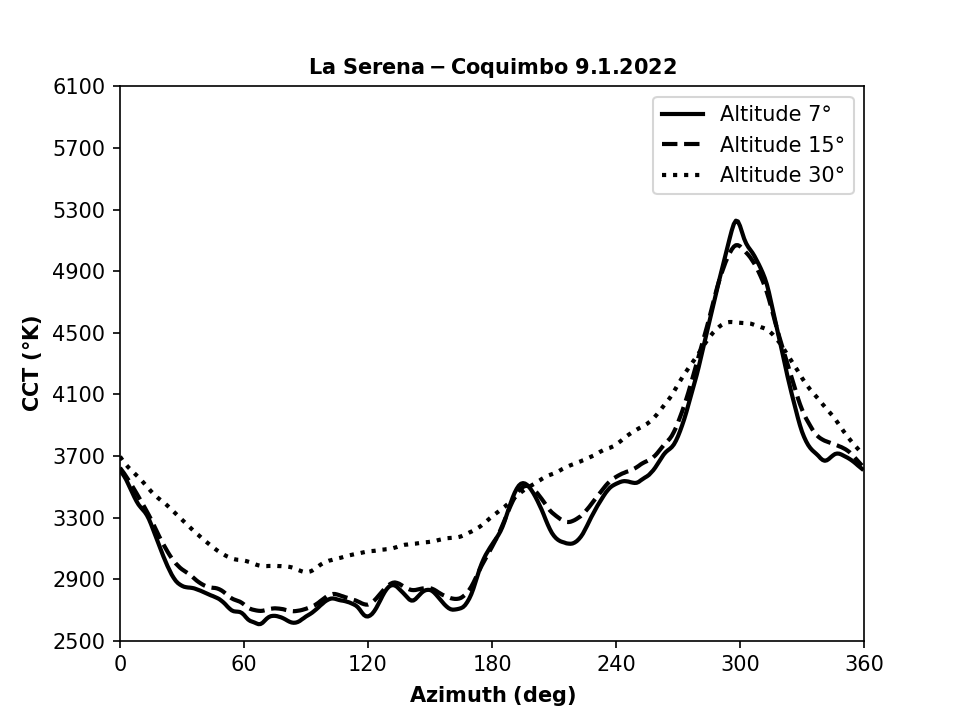}
    \includegraphics[width=0.49\linewidth]{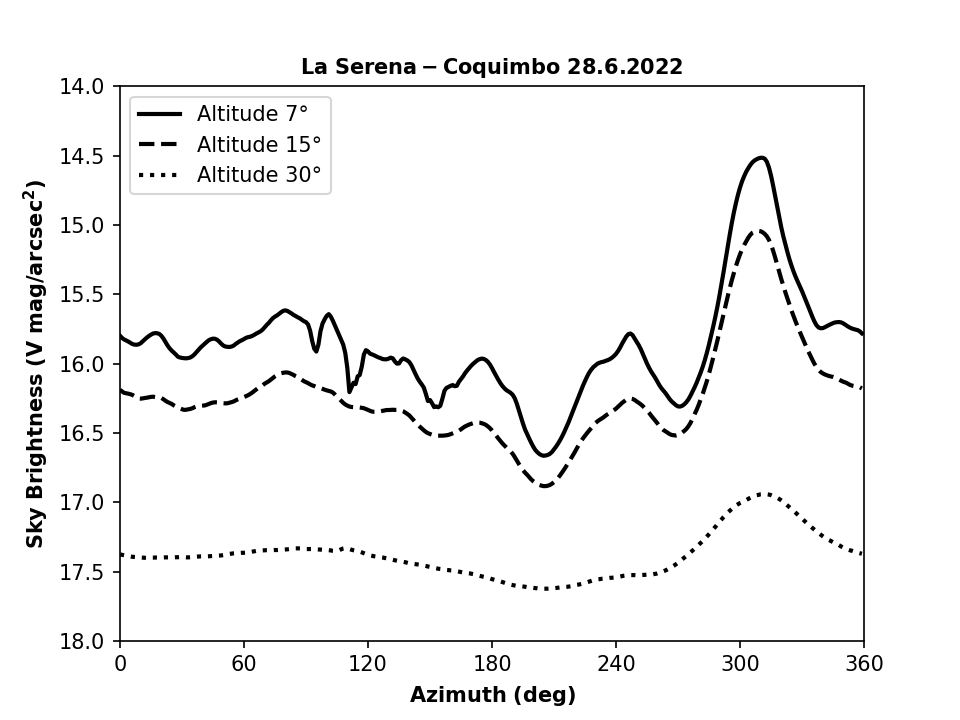}
    \includegraphics[width=0.49\linewidth]{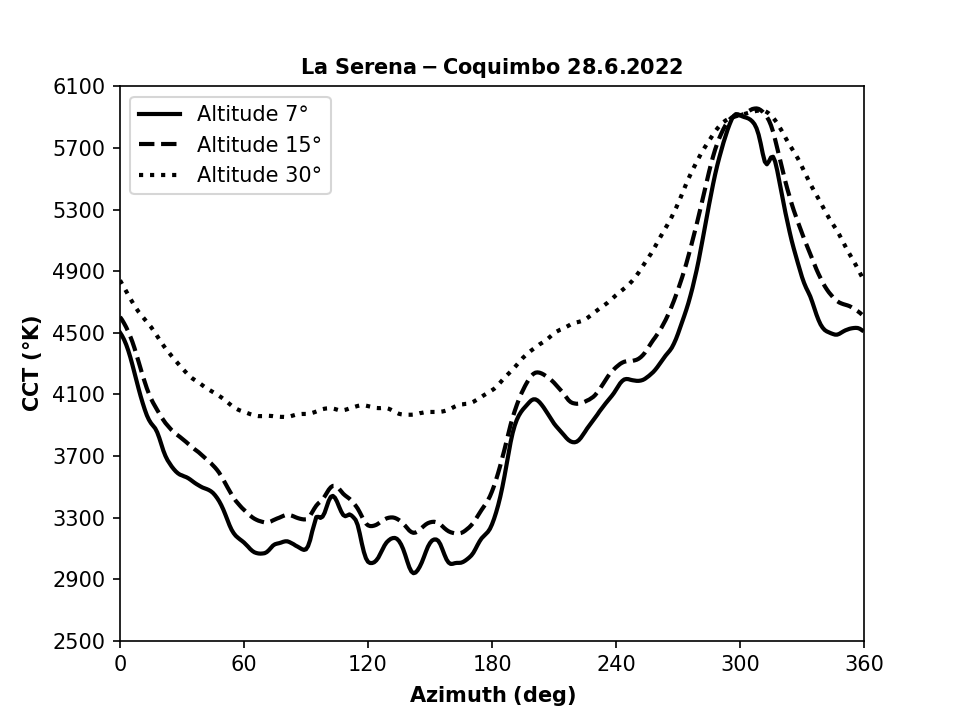}
    
    \caption{LS-CQ NSB/CCT profiles averaged within 1$\degree$-wide rings centered at 7$\degree$, 15$\degree$, and 30$\degree$ elevation, as a function of azimuth (see also Table \ref{tab:areas}).}
    \label{fig:ita_azring}
\end{figure*}

The SQC RGB images and corresponding NSB/CCT maps from our urban survey are shown in Figure  \ref{fig:sqc_ita}. The RGB images (left panels) reveal that the MK is completely lost in the artificial glare, and even second magnitude stars are barely discernible. The sky is so bright that a different color scale has to be adopted in order to maintain readability in the SQC maps (middle and right panels). Analysis of the azimuth profiles (Figure  \ref{fig:ita_azring}) confirms a uniformly bright sky due to artificial lighting, with one direction even exhibiting an unnatural whitish hue.

\begin{figure*}
    \centering
    \includegraphics[width=0.49\linewidth]{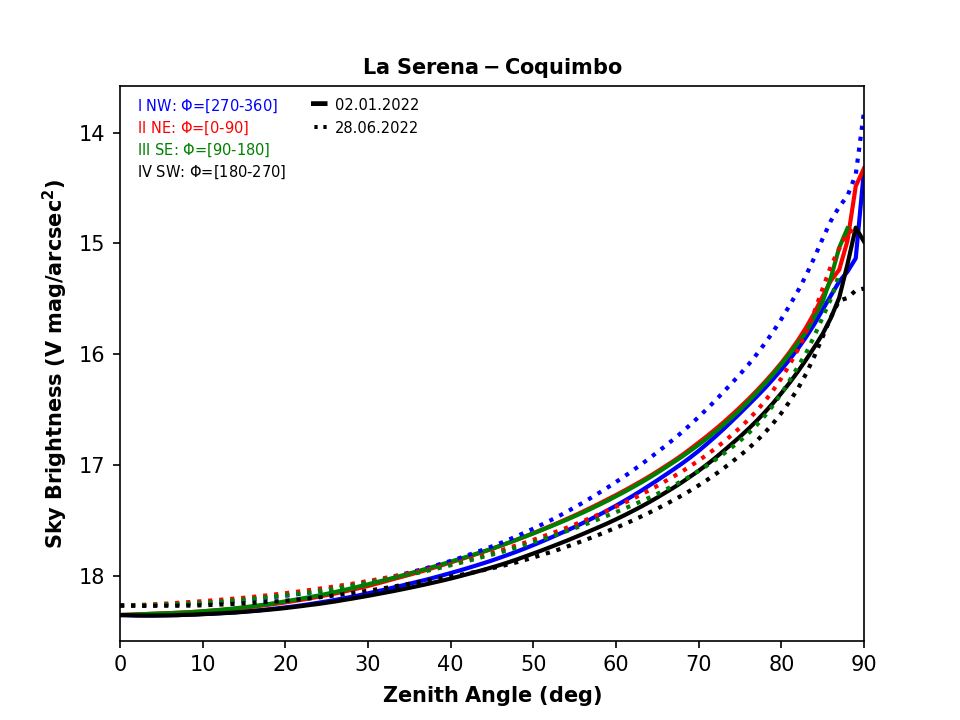}
    \includegraphics[width=0.49\linewidth]{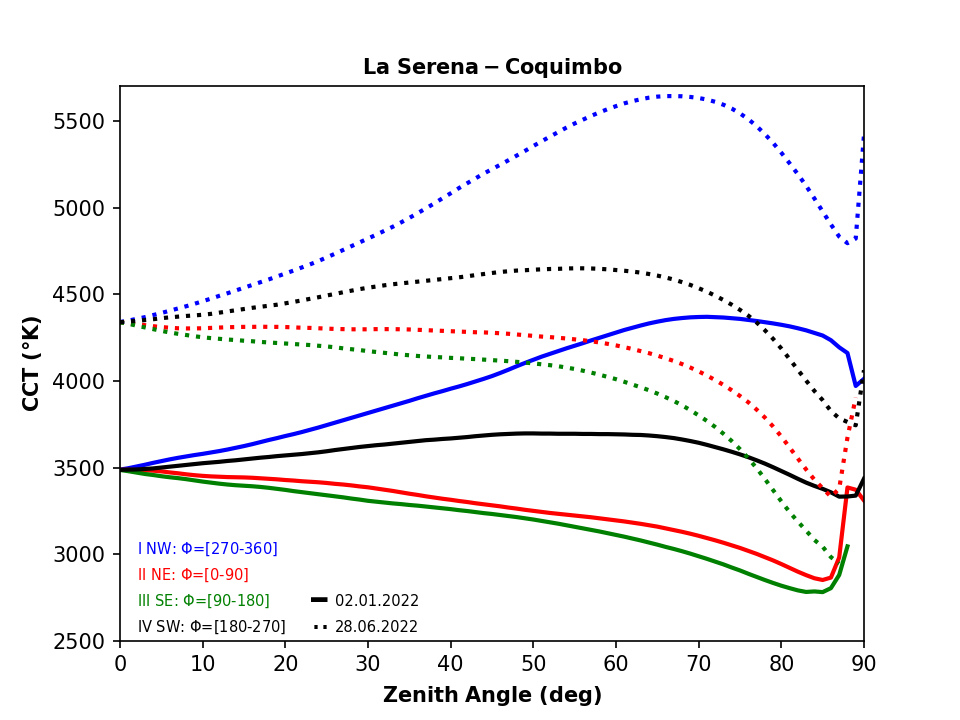}
    \caption{Same as Figure  \ref{fig:vertplot_fjnp}, but for LS-CQ on 2022 Jan 2 (solid lines) and 2022 Jun 28 (dotted lines). For further details refer to Sect. \ref{sec:ls}, Figure  \ref{fig:sqc_ita}, and Table \ref{tab:areas}.}
    \label{fig:vertplot_ita}
\end{figure*}

Under this heavily contaminated sky, the very same concept of ALAN-free direction loses its significance. For our analysis we therefore divided the sky into four cardinal quadrants (Figure  \ref{fig:vertplot_ita}). Except for a few degrees above the horizon, where the different lighting installations can still be individually discerned, the NSB profiles of the four quadrants overlap each other almost perfectly. This is a direct consequence of being virtually immersed in a homogeneously lit ALAN bubble. During our first visit in January, the primarily residential area towards the eastern horizon was dominated by low-CCT HPS sources still commonly used for public street lighting; while the western horizon already exhibited high CCT values predominantly from white-blue LEDs installed at private facilities such as amusement parks and commercial malls.

However, it is the comparison between the first and second visit (which are only one semester apart) to be truly dramatic. In June, we encountered slightly higher brightness level but, more significantly, the sky color had changed so much as to become discomforting to the naked eye in certain directions. The higher CCT values that in January were confined to a large but still delimited azimuth range, in June had expanded from multiple directions, covering almost the entire sky. Under these conditions, where even the darkest point in the sky is brighter than the brightest points recorded at all previous sites (cfr. Table \ref{tab:minmax}), and the brightest point approaches the boundary between mesopic and photopic vision  (\citealt{green2022growing}; \citealt{stockman2006}), only the Moon, the brightest planets and less than 1\% of stars (\citealt{kyba2023citizen}; \citealt{cinzano2020toward}) remain visible from the capital of the \textit{Región Estrella}.

\section{Discussion}\label{sec:discussion}
In the preceding section, we have outlined the most significant findings achieved by this first survey at each monitoring location. In this concluding section, we discuss them in a more general context by comparing the quality of the night sky in the Coquimbo Region with similar sites worldwide, as well as with state-of-the-art numerical models. \\

\subsection{Photometric indicators for cross-site comparisons}\label{subsec:indicator}
\begin{figure*}
    \centering
    \includegraphics[width=0.4\linewidth]{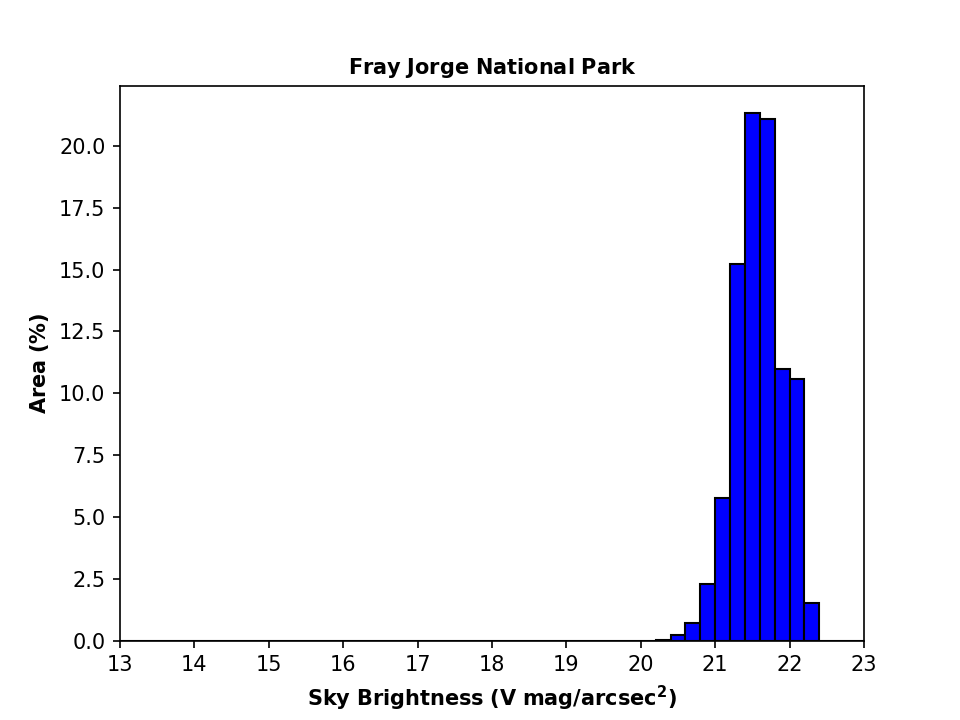}
    \includegraphics[width=0.4\linewidth]{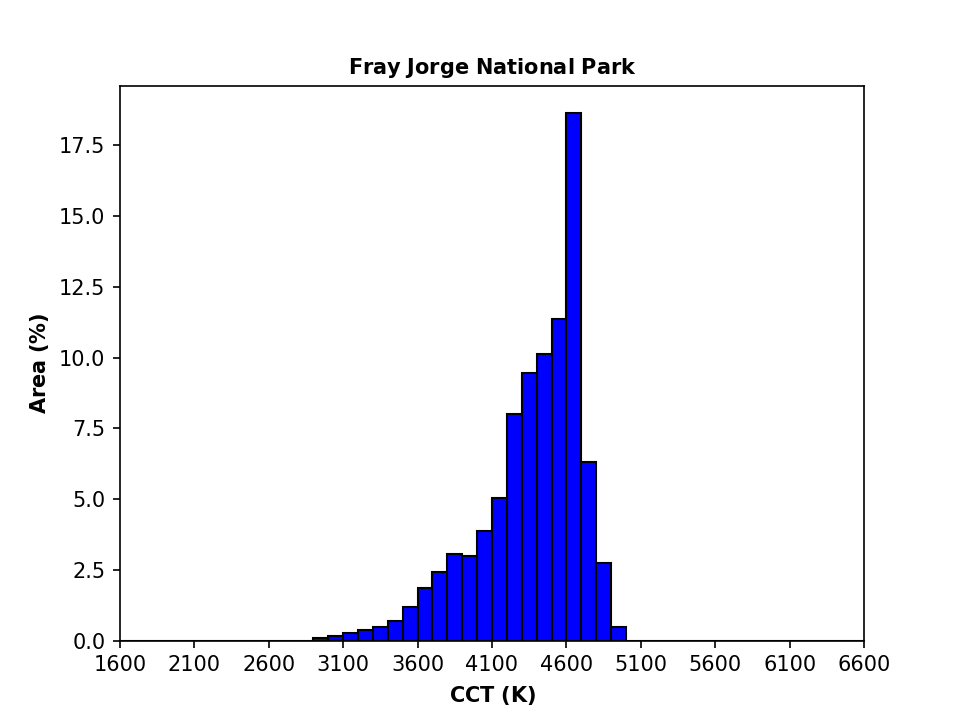}
    \includegraphics[width=0.4\linewidth]{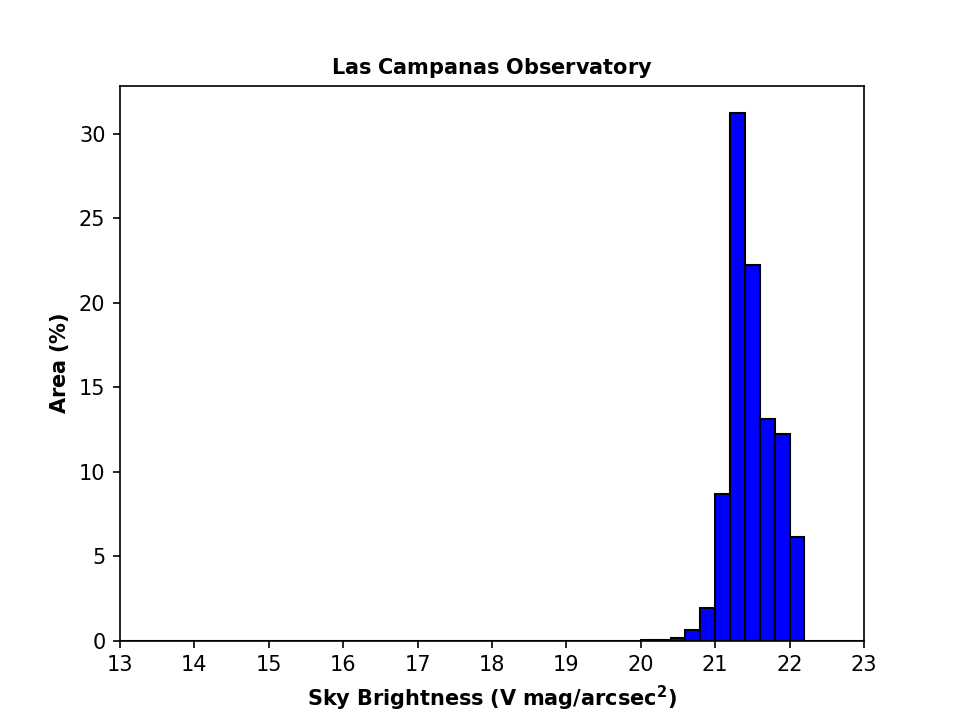}
    \includegraphics[width=0.4\linewidth]{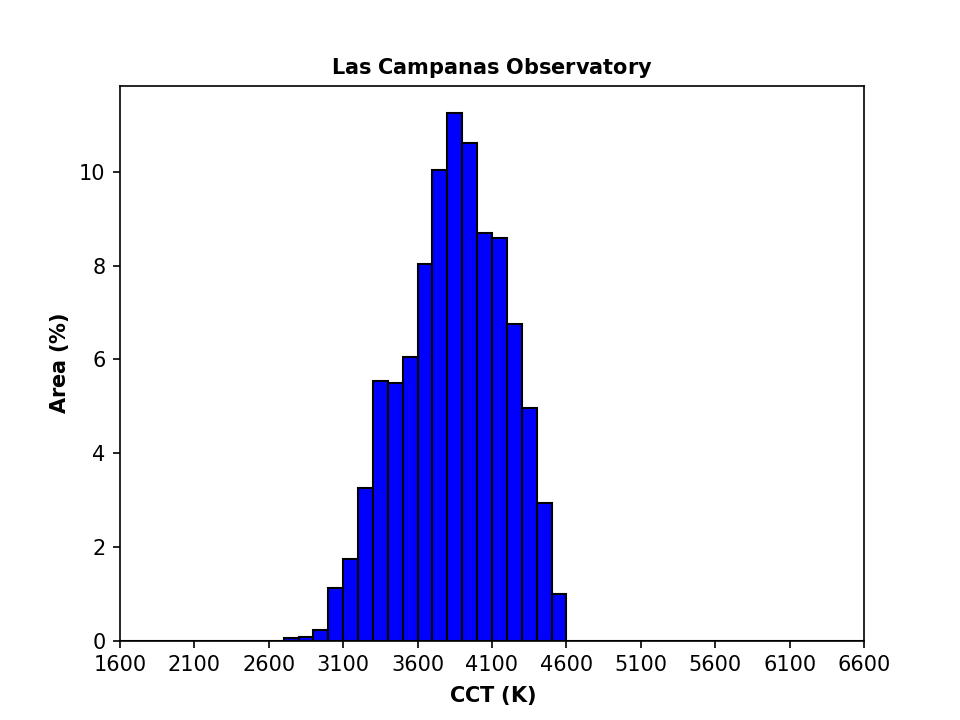}
    \includegraphics[width=0.4\linewidth]{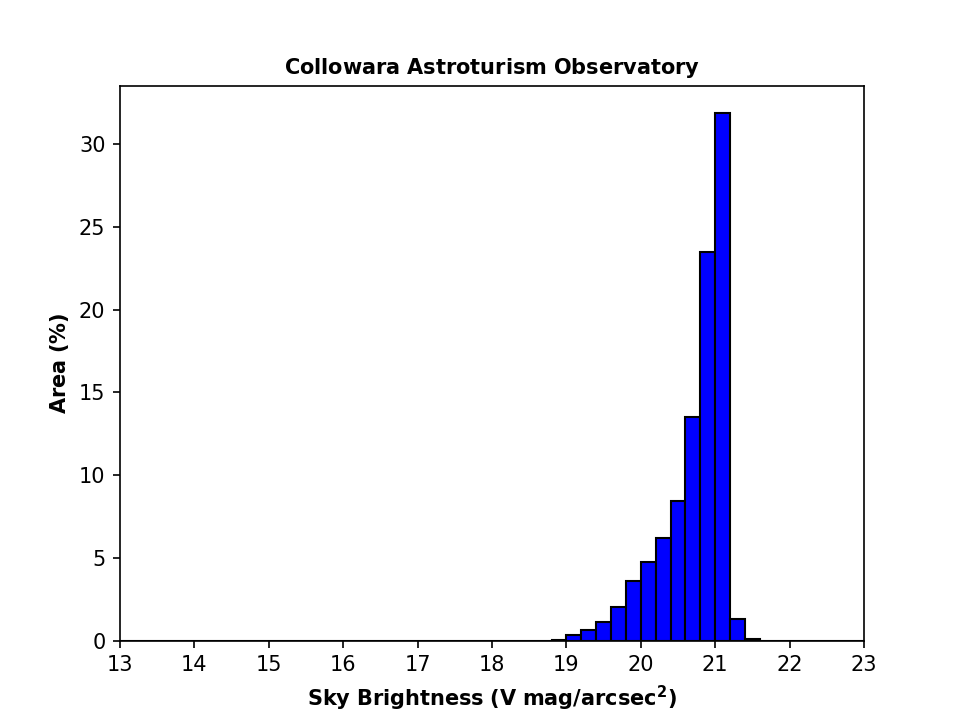}
    \includegraphics[width=0.4\linewidth]{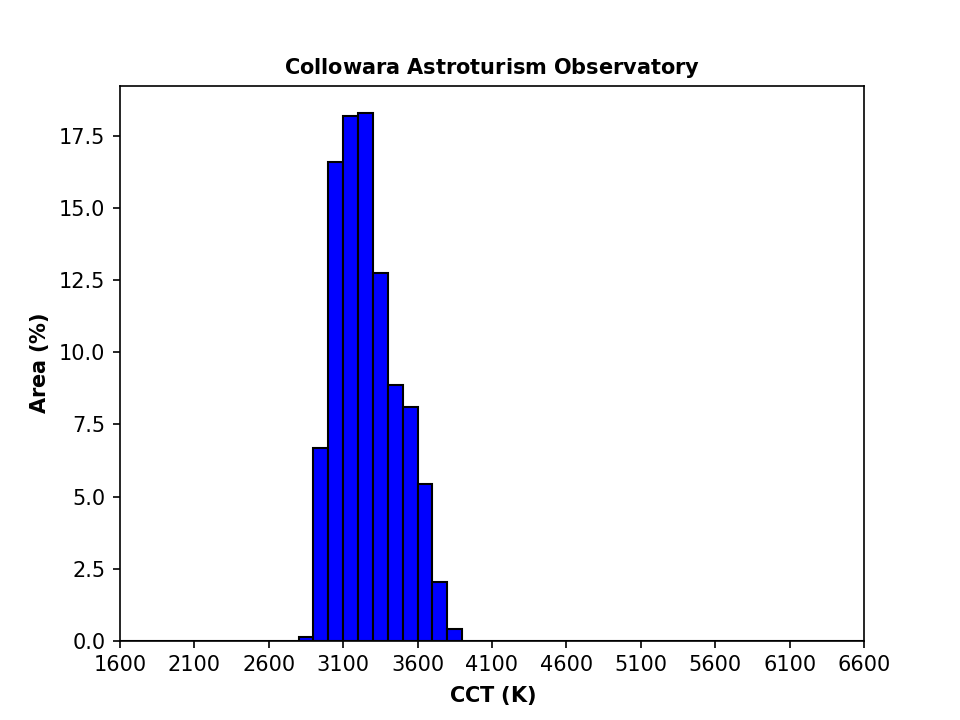}
    \includegraphics[width=0.4\linewidth]{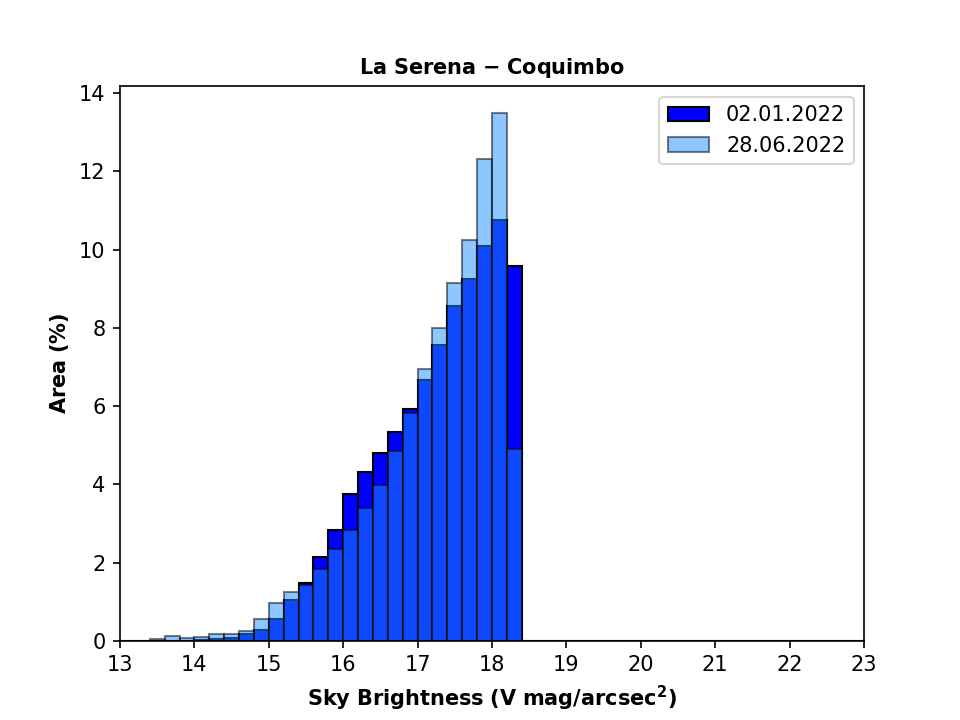}
    \includegraphics[width=0.4\linewidth]{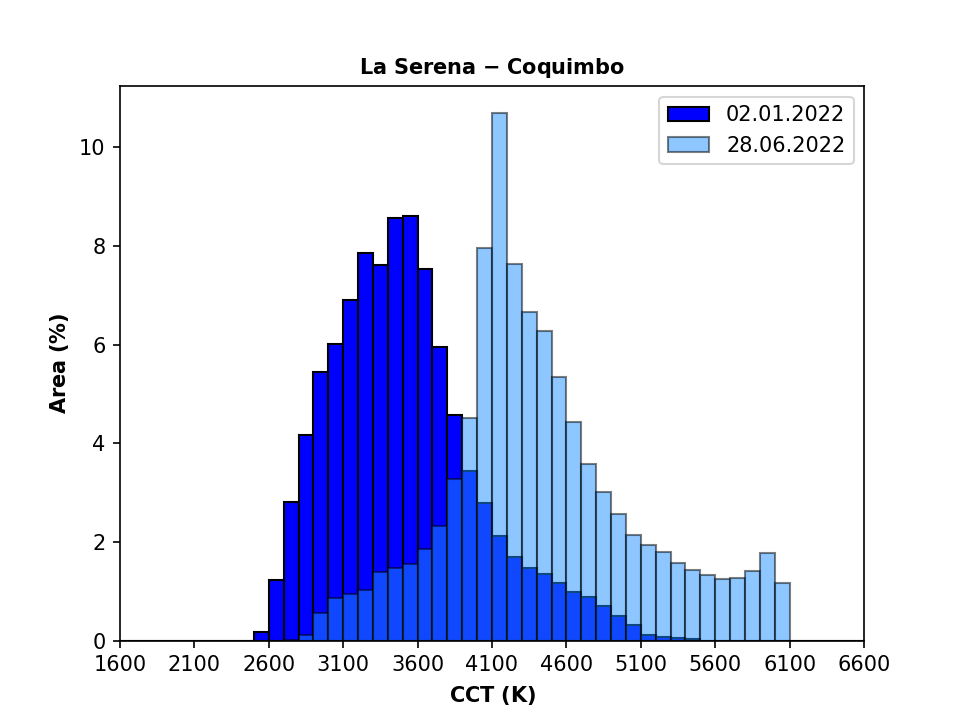}
    \caption{The histograms represent the distribution of all-sky NSB (left column) and CCT (right columns) values for each of the four measurement sites. The bin width is set at 0.2 mag/arcsec$^2$ and 100 K, respectively.}
    \label{fig:histograms}
\end{figure*}

For comparative studies across different locations, a wealth of photometric indicators of visual night sky quality have been suggested in literature (\citealt{deverchere2022towards}; \citealt{falchi2021computing}; \citealt{duriscoe2016photometric}). Each indicator has its advantages and disadvantages. For instance, the brightest section of the sky is considered significant for biological studies, including human health, while the darkest one serves as a reliable proxy for sky quality since it represents the celestial field least affected by light pollution, and therefore closest to natural conditions. Although zenith values are widely reported in the literature, their suitability as a general indicator of sky quality has been questioned, particularly for areas with low-level light pollution (\citealt{kollath2023natural}; \citealt{falchi2023light}). At the same time, the average NSB in the 30$\degree$ altitude ring is considered a reliable index for astronomical research, given that most professional astronomical observations are conducted above airmass 2. The most objective indicator to describe overall sky quality is however agreed to be the all-sky averaged NSB value (\citealt{falchi2023light}; \citealt{duriscoe2016photometric}). In Table  \ref{tab:minmax}, we provide the alt-az coordinates and corresponding NSB/CCT values for the darkest and brightest points in the sky, as well as at zenith (i.e., averaged over a circle of 5$\degree$ radius), within a 1$\degree$ wide ring centered at 30$\degree$ elevation, and across the entire celestial hemisphere (i.e., all-sky) for each site during our observations. Figure   \ref{fig:histograms} displays histograms of the all-sky NSB/CCT distributions, suitable for quickly identifying  at any given location the portion of celestial hemisphere above or below a specific reference threshold.\\

\begin{deluxetable}{ccccccccccccccc}
\caption{Alt-az coordinates and corresponding NSB (in V mag/arcsec$^2$) and CCT (in K) values of the darkest and brightest points in the sky at the time of our observations. Also shown are the averaged values at zenith, at 30$\degree$ fixed elevation, and over the entire celestial hemisphere.}
\label{tab:minmax}
\tablehead{
\colhead{Site} & \multicolumn{4}{c}{Brightest Point$^a$} & \multicolumn{4}{c}{Darkest Point$^a$} & \multicolumn{2}{c}{Zenith$^b$} & \multicolumn{2}{c}{30$\degree^c$} & \multicolumn{2}{c}{All-Sky} \\ \cline{2-15} 
\colhead{acronym} & \colhead{AZ} & \colhead{ZA} & \colhead{NSB} & \colhead{CCT} & \colhead{AZ} & \colhead{ZA} & \colhead{NSB} & \colhead{CCT} & \colhead{NSB} & \colhead{CCT} & \colhead{NSB} & \colhead{CCT} & \colhead{NSB} & \colhead{CCT}
}
\startdata
FJNP & 86 & 86 & 20.33 & 3325 & 216 & 25 & 22.25 & 4416 & 22.05 & 4926 & 21.57 & 4452 & 21.55 & 4311 \\
LCO & 208 & 90 & 19.96 & 2456 & 223 & 24 & 22.08 & 4004 & 22.02 & 4233 & 21.44 & 3978 & 21.45 & 3836 \\
CAO & 317 & 90 & 18.95 & 3042 & 145 & 57 & 21.19 & 3476 & 21.09 & 3485 & 20.77 & 3255 & 20.64 & 3250 \\
LS-CQ 1$^{st}$ & 304 & 87 & 14.29 & 4978 & 264 & 5 & 18.36 & 3533 & 18.34 & 3484 & 17.35 & 3466 & 16.90 & 3351 \\
LS-CQ 2$^{nd}$ & 309 & 86 & 13.99 & 5881 & 241 & 8 & 18.27 & 4398 & 18.26 & 4352 & 17.37 & 4594 & 16.81 & 4234 \\ 
\enddata
\tablecomments{$^a$ Average value within 1 deg$^2$; $^b$ Average value for zenith angle ZA$<$5$\degree$; $^c$ Average value at $30\degree$ elevation, in a $1\degree$ wide ring (i.e., for $59.5\degree<$ZA$<60.5\degree$).}
\end{deluxetable}

\subsection{Cross-site comparisons}\label{subsec:sites}
FJNP. Despite the asymmetry that sees at FJNP the western horizon (when considering only ZA$\geq$60°) 28\% brighter than the eastern one, the reference photometric indicators presented in Table \ref{tab:minmax} indicate that FJNP is one of the darkest measured locations on Earth\footnote{Along with few other natural parks located in continental US, such as Banner Creek Summit and Galena Lodge in Idaho's Sawtooth National Recreation Area -- \texttt{https://www.nps.gov/subjects/nightskies/skymap.htm}} \citep{alarcon2021natural}. The further comparison with state-of-the-art numerical models (Figure \ref{fig:gambons} and Table \ref{tab:gambons}) confirms this conclusion by estimating that ALAN contributes only $\sim$4\% to FJNP overall NSB.
This evidence, coupled with the exceptional ecological importance of FJNP as a one-of-a-kind ecosystem, highlights the urgent need for coordinated conservation efforts that go beyond the existing measures in place. While the importance of FJNP as a UNESCO Biosphere Reserve is widely acknowledged and the access to the site is regulated accordingly, the same level of active protection has not been extended to its status as a Starlight Reserve. To the best of our knowledge, in fact, there are currently no concrete actions focused on preserving the pristine quality of its night sky, nor ongoing monitoring programs able to provide timely alerts in the event of any significant deviations of the relevant variables.\\

LCO. The increasing evidence in the VIIRS yearly data releases of ALAN spots associated with the renewed lighting system of the panamerican highway has raised concerns regarding its potential impact on the night sky quality of nearby professional observatories, particularly LCO and ESO La Silla (\citealt{blanc2019}; \citealt{falchi2023light}). \cite{falchi2023light} estimated that ``the lights from Ruta 5 alone, within a 40 km distance from the observatory, contribute to over 50\% of both the artificial zenith radiance and the average radiance at 30$\degree$'' at LCO. However, the recorded SQC data towards Ruta 5 approximately indicate the same NSB levels as those in the antipodal direction. The lack of spectral information on a night characterized by high airglow activity prevents us from accurately determining the specific contributions of each ALAN source, nor to weight it against a measured (i.e., not assumed \textit{a priori}) natural reference level. We are therefore not in the position to confirm or dismiss Falchi's statement. As we have emphasized throughout this paper, determining a reliable ratio of artificial to natural NSB without dedicated spectroscopy is extremely challenging. On the other side, it is also worth mentioning that the convolution of a point spread function with the geographical distribution of light sources, although useful for mapping ALAN radiance over large areas, may still suffer from considerable uncertainties and could overestimate the artificial contribution by up to 50\% \citep{simoneau2021point}. Further data collection and collaborative studies are thus necessary to improve our current understanding of this situation. Nonetheless, the analysis of various metrics that includes the NSB value of the darkest point in the sky (V=22.08 mag/arcsec$^2$) and at zenith (V=22.02 mag/arcsec$^2$) as shown in Table \ref{tab:minmax}, and the comparison with numerical simulations (for which ALAN contributes only $\sim$11\% to the overall NSB - Figure \ref{fig:gambons} and Table \ref{tab:gambons}) reaffirm that LCO is one of the best locations worldwide for conducting cutting-edge astronomical research. If it will remain so also in the upcoming ELT era, it only depends on our \textit{present} ability to defend it against the threat of the surrounding ALAN sources, first and foremost of the rapidly expanding urban areas.\\

CAO. Due to the geographical distribution of ALAN sources surrounding CAO, the western celestial hemisphere is heavily affected by skyglow up to at least 50$\degree$ elevation. In contrast, the eastern hemisphere appears noticeably darker, with the darkest region in the sky located only 30$\degree$ above the horizon, pretty close to the brightest natural feature represented in this case by the MK bulge. Our SQC data allows us to quantify this asymmetry: the western hemisphere is approximately 70\% brighter than the eastern one, and this percentage increases to about 130\% when considering only the first 30$\degree$ above the horizon. The artificial contribution to NSB overwhelms the natural one even at the zenith. 
It is thought-provoking to realize that the night sky quality indicators at CAO are comparable with the ones of Flagstaff, Arizona\footnote{SQC images recorded in Flagstaff on August 16, 2017 are publicly available at \texttt{https://lightpollutionmap.info}.}, highlighting the fact that a city with $\sim$11,000 inhabitants (Andacollo) is plagued by light pollution levels that are similar to those of a much larger urban area with nearly 140,000 residents like Flagstaff. On one hand, this comparison praises the long-term measures taken by Flagstaff local administration to regulate ALAN, which was awarded in 2001 the title of world’s first Dark Sky Community\footnote{\texttt{https://www.flagstaff.az.gov/3799/Dark-Sky-Community}}; it also proves that well-coordinated efforts to protect the night sky do not necessarily hinder any economic or demographic growth, nor represent a specific threat to public safety. On the other hand, it highlights that there is still much work that can (and should) be done in Chile to keep the light pollution phenomenon under control. \\     

LS-CQ. Probably one of the most significant findings of this study is that LS-CQ consistently emerges as the most invasive source of artificial light at the horizon of all of our observation sites, regardless of whether we sit high in the Andes near the northern border of the Coquimbo region, or deep in a ocean-side forest near its southern extreme. It is therefore imperative to prioritize efforts in controlling light pollution over this urban area. \\
As we have previously discussed (Section \ref{sec:ls}), the NSB/CCT values in LS-CQ indicate a rapid degradation of its night sky, which can be primarily attributed to the wide-spreading use of white-blue LED lighting. Interestingly enough, we observationally confirm the trends suggested by \cite{lughi2014} models, namely that blue-rich LED sources \textit{``produce a dramatically greater sky brightness than yellow-rich HPS sources [...] when matched lumen-for-lumen and observed nearby"}, but at the same time their contribution decreases faster with distance because of a higher scattering efficiency. In other words, the copious LED contribution from LS-CQ causing the high CCT values of Figure  \ref{fig:sqc_ita} is largely scattered out by the time it is recorded at the other sites, explaining the CCT dips to $\sim$ 3000 K in the direction of LS-CQ of Figures  \ref{fig:fjnp_azring}, \ref{fig:lco_azring},\ref{fig:cao_azring}.

To truly grasp the severity of the situation, it is enlightening to compare the sky quality indicators of LS-CQ with those of Padova, a city in northern Italy of similar size and at approximately the same elevation above sea level, but located in one of the most densely populated (and light-polluted -- \citealt{falchi2016new}) areas of Europe.
Publicly available SQC images from downtown Padova captured on November 11, 2019 reveal virtually identical NSB within a few tenths' magnitude\footnote{Padova's darkest spot in the sky: 18.28 mag/arcsec$^2$; NSB within the 30$\degree$ ring: 17.46 mag/arcsec$^2$; all-sky radiance: 17.49 mag/arcsec$^2$. Compare these values with the last two rows of Table \ref{tab:minmax}.} and even lower CCT values compared to those measured in LS-CQ during our two visits in 2022. It is therefore extremely concerning to realize that if one can still refer to a general darkness of the Chilean skies, and La Serena be praised as the capital of the \textit{Región Estrella}, it is not for the results of an active and effective control of light pollution, but it is fortuitously due to the urban sparse distribution of a low-populated country with a unique geography.

\begin{figure}
    \centering
    \includegraphics[width=0.75\linewidth]{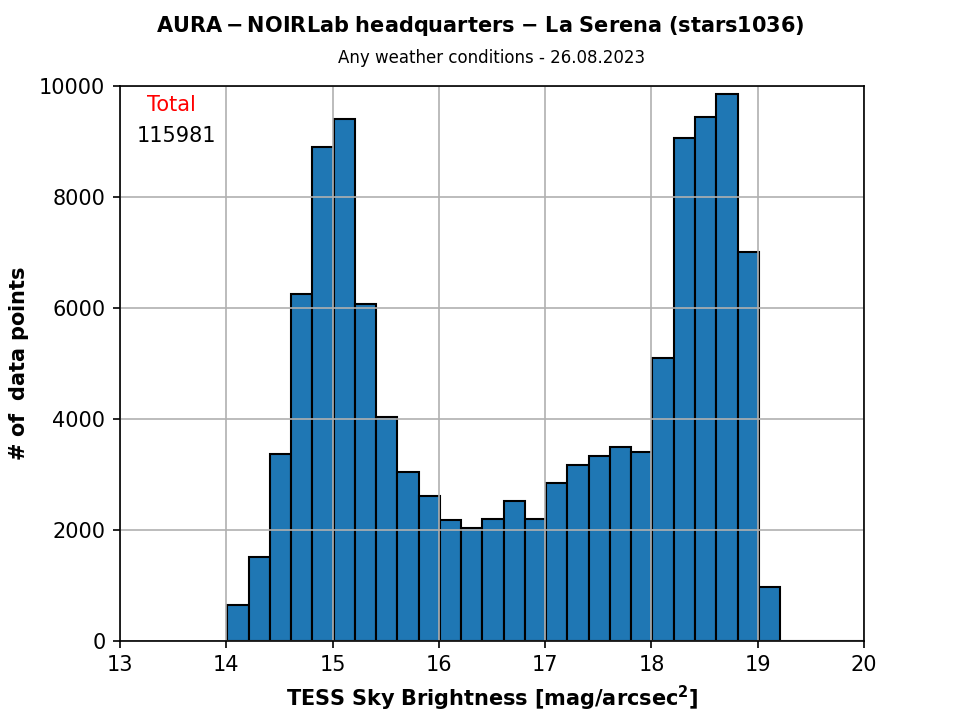}
    \includegraphics[width=0.75\linewidth]{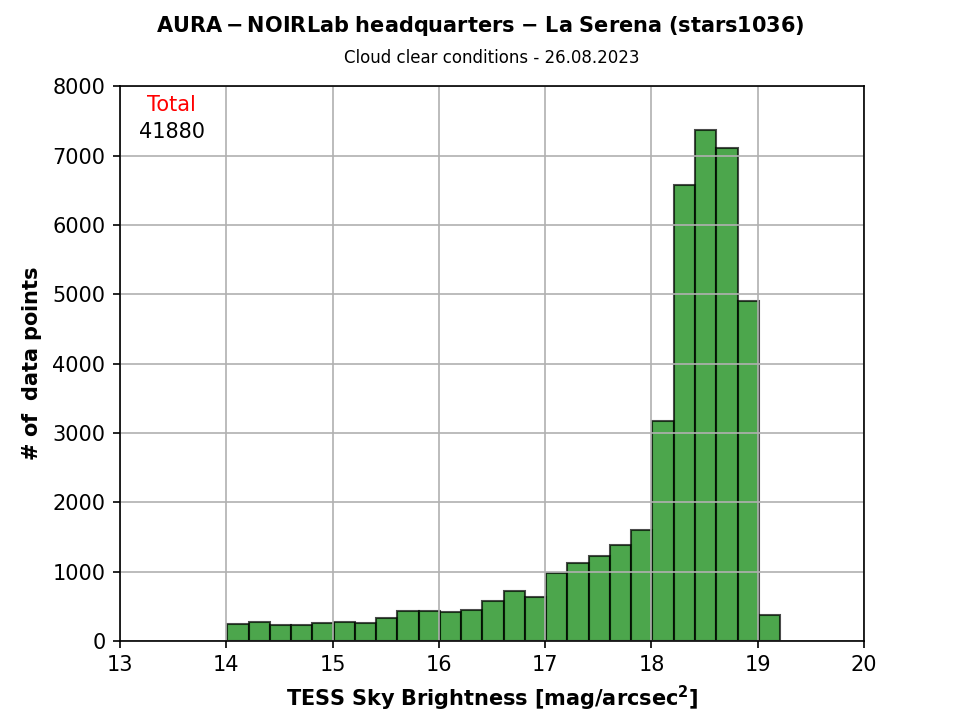}
    \caption{TESS-W distribution of zenith NSB in LS-CQ.  The histogram reflects over 115,000 TESS-W measurements taken in the $\sim$160 days leading up to 2023.08.26. The top panel collects measurements obtained under any weather conditions, while the bottom panel represents only measurements obtained under clear sky conditions. Clear conditions were observed approximately 37\% of the time, partially cloudy conditions were observed 23\% of the time, and completely overcast conditions were observed 40\% of the time.}
    \label{fig:tess_histograms}
\end{figure}

To monitor with the deserved pace the rapid evolution of ALAN in LS-CQ, we have recently installed a Telescope Encoder and Sky Sensor (TESS-W, \citealt{zamorano2016stars4all}) in the AURA-NOIRLab headquarters in La Serena. This TESS-W (stars1036) has been continuously monitoring the night sky at zenith since March 2023, and its NSB measurements are summarized in Figure  \ref{fig:tess_histograms}. They reveal a clear bimodal distribution and are consistent with the SQC-derived zenith values recorded under clear-cloud conditions during our two 2022 visits (Table \ref{tab:minmax}). It is well-known that in urban environments variable cloud coverage can amplify ALAN \citep{kyba2011cloud}: in LS-CQ our TESS data suggest that completely overcast nights appear roughly 25 times brighter than moonless, cloud-clear nights.
Thanks to publicly funded programs, we are currently assembling a network composed of $\sim$40 TESS stations that will be monitoring professional and astrotourism observatories, natural reserves, and rural schools across the entire \textit{Región Estrella} (Jaque Arancibia et al., in preparation).

\subsection{Comparison with GAMBONS numerical models}\label{subsec:gambons}
\begin{figure*}
    \centering
    \includegraphics[width=0.95\linewidth]{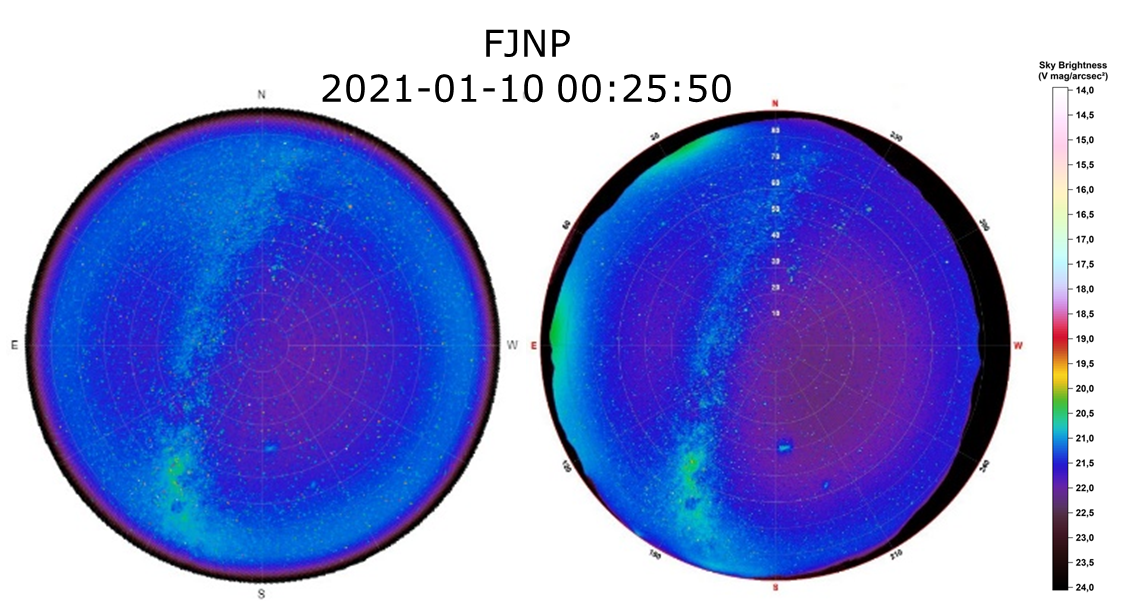}
    \includegraphics[width=0.95\linewidth]{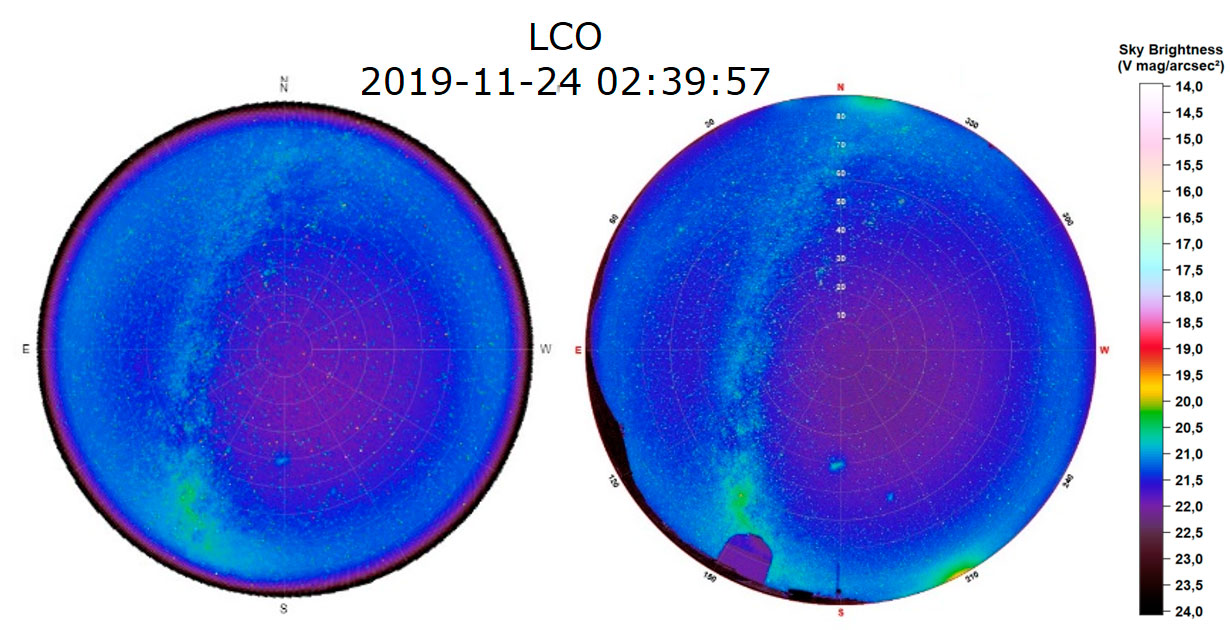}
    \caption{NSB maps according to the GAMBONS numerical models (left panels) vs. SQC-based observations (right panels) for FJNP (upper row) and LCO (bottom row). Please refer to Sect. \ref{sec:discussion} and Table \ref{tab:gambons} for further details.}
    \label{fig:gambons}
\end{figure*}

\begin{figure*}
    \centering
    \includegraphics[width=0.95\linewidth]{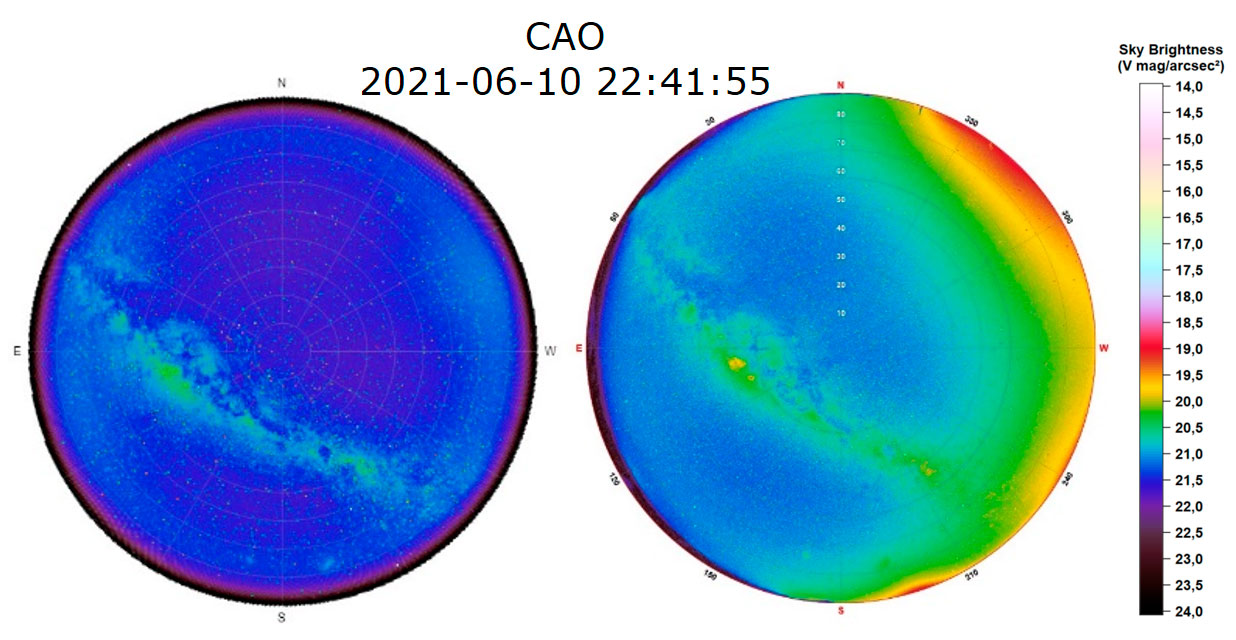}
    \includegraphics[width=0.95\linewidth]{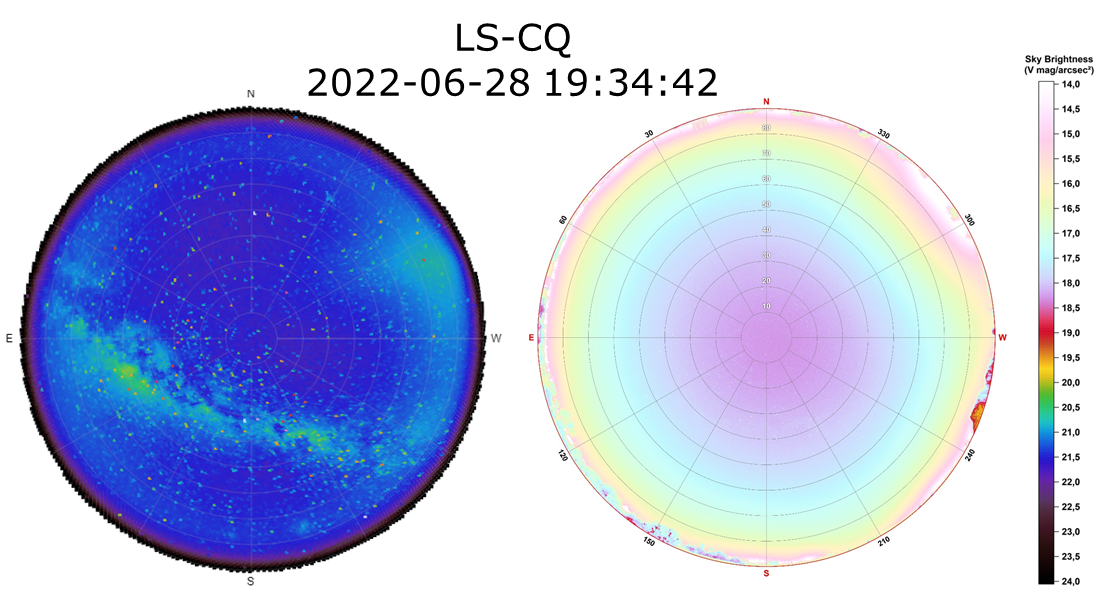}
    \caption{As for Figure  \ref{fig:gambons} but for CAO and LS-CQ. Refer to Sect. \ref{sec:discussion} and Table \ref{tab:gambons} for further details.}
    \label{fig:gambons2}
\end{figure*}

Figures \ref{fig:gambons} and \ref{fig:gambons2}, along with Table \ref{tab:gambons}, provide a direct comparison between our SQC data and the GAMBONS numerical models (\citealt{masana2021multiband, masana2022enhanced}). GAMBONS offers a user-friendly web interface\footnote{\texttt{https://gambons.fqa.ub.edu/index.html}} that allows users to generate realistic models of the sky's natural brightness in cloudless and moonless nights for any given time and location on Earth. These models take separately into account the brightness contributions of stars from Gaia ED3 and Hipparcos catalogs, diffuse galactic and extragalactic light, zodiacal light, airglow, and atmospheric attenuation and scattering effects.
The models have been run using default airglow and aerosol parameters and can be considered a good first approximation to the typical conditions of our monitoring sites. However, it is important to note that the models available on the public web interface adopt a simplified approach to evaluate the impact of atmospheric scattering in order to expedite computation time. These simplified simulations are known to overestimate the natural NSB near the horizon and underestimate it at zenith by up to 0.1 mag/arcsec$^2$ \citep{masana2022enhanced}. Taking into account both the modeling accuracy and the precision of our data, the agreement between the natural (as provided by GAMBONS) and observed (i.e., natural + artificial, as provided by the SQC) zenith NSB for the most pristine site is remarkable. Within this level of confidence, it becomes also possible to quantify the artificial contribution to the overall radiance of the night sky (last column of Table \ref{tab:gambons}). It is worth noticing that if FJNP is confirmed to currently have a negligible level of light pollution, and LCO is still in the list of excellent sites for conducting cutting-edge astronomical observations, the skyglow at CAO has already erased nearly half of the stars that were visible by the naked eye until a few human generations ago. ALAN in LS-CQ has already killed the natural starlight, resulting in nights virtually devoid of stars. \\

\begin{deluxetable}{ccccccc}
  \caption{Comparison between GAMBONS numerical models (GMB) and observed (SQC) NSB at zenith and all-sky, in V mag/arcsec$^2$. The last column indicates the artificial contribution to the overall (i.e., observed $=$ natural $+$ artificial) radiance of the night sky. See Sect. \ref{sec:discussion} for further details.}
  \label{tab:gambons}
  \tablehead{
  \colhead{Site} & \multicolumn{3}{c}{Zenith$^a$} & \multicolumn{3}{c}{All-Sky} \\ \cline{2-7} 
  \colhead{acronym} & \colhead{GMB} & \colhead{SQC} & \colhead{$\Delta$m} & \colhead{GMB} & \colhead{SQC} & \colhead{ALAN \%}
  }
  \startdata 
   FJNP & 21.83 & 21.90 & -0.07 & 21.52 & 21.48 & 4.0 \\
   LCO & 21.78 & 21.95 & -0.17 & 21.52 & 21.39 & 11.3 \\
   CAO & 21.52 & 21.05 & 0.47 & 21.51 & 20.61 & 56.6 \\
   LS-CQ 1$^{st}$ & 21.80 & 18.35 & 3.45 & 21.53 & 16.89 & 98.6 \\
   LS-CQ 2$^{nd}$ & 21.66 & 18.26 & 3.40 & 21.47 & 16.81 & 98.6 \\ 
\enddata
\tablecomments{$^a$ Average NSB for ZA$<$5$\degree$.}
\end{deluxetable}

\section{Concluding Remarks}\label{sec:conclusion}
We want to close this work by further calling the attention on the potentially confusing similarity between the high CCT values recorded at both FJNP and LS-CQ (Figure  \ref{fig:histograms}). In Section \ref{sec:intro} we had briefly discussed the ALAN SPD reshaping that it is taking place because of the so-called \textit{LED revolution}: among the various detrimental effects it is at the origin of, there is in fact also the increasing evidence that most of the photometric technologies and techniques that have so far been used to monitor ALAN worldwide may not able to keep pace with the massive and fast spread of these cheap, and therefore widely accessible, lighting sources. This is mainly due to the diverging spectral emissivity of lighting sources on one side, and the spectral sensitivity range of the instruments devoted to the measurements of night sky quality on the other side \citep{kyba2023citizen}.

\cite{falchi2016new} already noted that the VIIRS DNB sensor spectral response ``leaves out the blue and violet parts of the visible spectrum [...] thus preventing a good control of the evolution of light pollution in this important spectral band, where the white LEDs now being installed have strong emissions." 
Even the more advanced DSLR all-sky cameras which are increasingly used in professional ALAN investigations  (e.g., \citealt{vandersteen2020quantifying}; \citealt{jechow2019dark}; \citealt{kollath2017night}; this same study) may face challenges in distinguishing between natural and artificial sources of NSB as soon as the blue-shifting of ALAN SPD becomes even more pronounced \citep{vandersteen2020quantifying, jechow2019dark, kollath2017night}. While incandescent and HPS lamps can be differentiated from the natural sky emission because of their different CCT, the contamination from typical LED sources that emit 70\% of light in the blue-green (400-480 nm) range becomes basically indistinguishable to the eyes of SQC and similar devices, as we have also reported in our study. This highlights the need for improved measurement techniques and analysis methods to effectively separate natural and artificial sources of NSB in the face of increasing LED lighting.

Significant progress in the quantitative monitoring of ALAN could be achieved through spectroscopy, which so far has primarily been conducted at professional astronomical observatories \citep{patat2008dancing}. This approach allows for a much direct and trusty identification of natural and artificial contributions to the sky brightness level, also enabling the precise tracking of its complex and multi-periodic variations over time. In collaboration with the University of Padova, we are currently working on the development of a portable spectrograph capable of capturing low-resolution (R$\sim$600), optical (350-750 nm)  flux-calibrated spectra of large portion ($\sim$30$\degree$) of the night sky at once. The first prototype is already collecting continuous data at the Agency for Environmental Protection of Veneto Region (ARPAV) headquarters in Padova, Italy. An improved version is planned to be installed near the 8.1 m Gemini South telescope on Cerro Pachón during semester 2024A, while a second, portable one, will be joining our SQC monitoring campaign across the Coquimbo Region and beyond.\\

All these scientific and instrumental projects, along with free-of-charge lighting consultancy initiatives and outreach campaigns\footnote{See, e.g., the educational videos of the Universidad de\\ La Serena web series \textit{\#IluminAconCiencia} on YouTube.}, are currently playing a crucial role in raising public awareness about the harmful impacts of light pollution in Chile. They help dispel misconceptions and promote a better understanding of appropriate lighting practices. The scientific evidence presented in this paper has also already offered valuable support to public committees at the municipal, regional, and national levels and has empowered them to formulate data-driven strategies to protect the darkness of the Chilean sky: a natural and cultural\footnote{\texttt{https://www3.astronomicalheritage.net}} heritage that is our scientific, social, and moral responsibility to defend and preserve for all future generations.

\begin{acknowledgments}
The authors would like to thank the referee for
helpful comments and suggestions that improved the manuscript.
The authors also express their gratitude to all those who provided logistical support during the visits to the observing sites: special thanks are extended to Solange Mondaca of COA and Carlos Slomp Burger of the Italian School of La Serena for granting and coordinating access to their respective facilities during the challenging period of the pandemic.
JPUT and RA acknowledge financial support of DIDULS/ULS through project PTE1953851. JPUT also acknowledges financial support of the National Agency for Research and Development (ANID) Scholarship Program/Doctorado Nacional/2021-21210732. MJA acknowledges the financial support of DIDULS/ULS through project PR2353855. DFO, MJA and RA acknowledge financial support of DIDULS/ULS through project PTE2053852.
\end{acknowledgments}


\bibliographystyle{aasjournal.bst}
\bibliography{alan_chile.blb}



\end{document}